\documentclass[lettersize,journal]{IEEEtran}
\usepackage{amsmath,amsfonts}
\usepackage{algorithmic}
\usepackage{algorithm}
\usepackage{array}
\usepackage[caption=false,font=normalsize,labelfont=sf,textfont=sf]{subfig}
\usepackage{textcomp}
\usepackage{stfloats}
\usepackage{url}
\usepackage{verbatim}
\usepackage{graphicx}
\usepackage{cite}
\usepackage{booktabs}
\usepackage{pifont}
\usepackage{colortbl}
\usepackage{color}
\usepackage{xcolor}
\usepackage{comment}
\definecolor{ikgreen}{RGB}{200,255,200}
\definecolor{ikredl}{RGB}{255,200,200}
\definecolor{ikred}{RGB}{255,100,100}
\newcolumntype{a}{>{\columncolor{ikgreen}}c}
\newcolumntype{b}{>{\columncolor{ikred}}c}
\newcolumntype{b}{>{\columncolor{ikredl}}c}
\def\code#1{\texttt{#1}}

\usepackage{enumitem}
\usepackage{subcaption}
\usepackage{makecell}
\usepackage{balance}
\usepackage{tabu, booktabs}
\AtBeginDocument{%
  \providecommand\BibTeX{{%
    \normalfont B\kern-0.5em{\scshape i\kern-0.25em b}\kern-0.8em\TeX}}}
\usepackage{titlesec}

\begin{document}

\title{Pro-ZD: A Transferable Graph Neural Network Approach for Proactive Zero-Day Threats Mitigation}

\author{\IEEEauthorblockN{
Nardine Basta\textsuperscript{\textsection},
Firas Ben Hmida\textsuperscript{\textsection}, Houssem Jmal\textsuperscript{\textsection},
Muhammad Ikram, \\ Mohamed Ali Kaafar, and
Andy Walker}}




\maketitle
\begingroup\renewcommand\thefootnote{\textsection}
\footnotetext{First authors, ordered alphabetically.}
\begingroup\renewcommand\thefootnote{}
\footnotetext{Nardine Basta, Muhammad Ikram, and Mohamed Ali Kaafar are with the Department of Computing, Macquarie University, NSW
2109, Australia}
\footnotetext{Houssem Jmal and Firas Ben Hmida are with the Signals and Systems
Department, Ecole Polytechnique de Tunisie, Carthage 2078, Tunisia.}
\footnotetext{Corresponding author: Nardine Basta, e-mail: nardine.basta@mq.edu.au.
}

\footnotetext{Andy Walker is with Ditno Inc., Sydney, NSW 2000, Australia.}
\begin{abstract}

In today's enterprise network landscape, the combination of perimeter and distributed firewall rules governs connectivity. To address challenges arising from increased traffic and diverse network architectures, organizations employ automated tools for firewall rule and access policy generation. Yet, effectively managing risks arising from dynamically generated policies, especially concerning critical asset exposure, remains a major challenge. This challenge is amplified by evolving network structures due to trends like remote users, bring-your-own devices, and cloud integration. This paper introduces a novel graph neural network model for identifying weighted shortest paths. The model aids in detecting network misconfigurations and high-risk connectivity paths that threaten critical assets, potentially exploited in zero-day attacks---cyber-attacks exploiting undisclosed vulnerabilities. The proposed Pro-ZD framework adopts a proactive approach, automatically fine-tuning firewall rules and access policies to address high-risk connections and prevent unauthorized access. Experimental results highlight the robustness and transferability of Pro-ZD, achieving over 95\% average accuracy in detecting high-risk connections.
\end{abstract}


\section{Introduction}

Firewalls play a crucial role in managing network traffic based on established security policies within an organization. Traditionally, these protective measures are stationed at the network perimeter to safeguard an organization's private network from external threats. The evolution of firewalls, particularly Next-Generation Firewalls (NGFW) like zero-trust (ZT) firewalls, has seen a shift towards distributed deployment offering protection to internal sections, such as data centers. 

As network traffic types and volumes continue to rise, automated tools become essential for the generation and deployment of firewall rules and network security policies to maintain high uptime. However, the automation of rule generation introduces the risk of errors, necessitating thorough validation to identify potential misconfigurations and high-risk network connections that may expose critical assets. 

The validation process is complex given the dynamic nature of contemporary network structures shaped by trends like remote users, bring-your-own devices, and cloud integration. The rise in malicious activities further exposes networks to new threats, elevating the risk of zero-day attacks \cite{crimes}. Stemming from unknown vulnerabilities, these attacks present a challenge due to the information asymmetry between attackers and defenders, making their exploits elusive.



While preventing zero-day attacks entirely is challenging, timely identification of high-risk connections and restricting access to critical network assets allows proactive mitigation of such threats \cite{AG_ZD}. Manual processes may not suffice for the necessary proactivity and real-time mitigation needed to address modern threats and dynamic network structures. Therefore, the adoption of an automated process for threat identification and mitigation becomes a crucial element in an organization's cybersecurity infrastructure.

Previous studies have explored the utilization of attack graphs for cyber risk assessment~\cite{byres, mcqueen, wang2013, AWAN2016}. These approaches investigate the dependencies between vulnerabilities and the network connectivity posture established by the firewall policies to identify potential attack paths. Other approaches relying on attack graphs focus on evaluating the risks associated with zero-day attacks. This involves predicting the quantity and locations of zero-day vulnerabilities, coupled with an analysis of the network connectivity structure to assess the likelihood of attackers exploiting these vulnerabilities~\cite{ZD1, k-zero, patrol, ZDATT}. 

This class of models relying on attack graphs and attack trees provides a systematic perspective on potential threat scenarios in networks leveraging the network connectivity structure. Nevertheless, their effectiveness is constrained by their inability to dynamically adapt to changes in the network connectivity structure. Any modification to the network structure requires the regeneration of the attack graph.

The literature has introduced various deep learning (DL) approaches~\cite{DL_cyber, deeplearningattacks, deeplearningSecurity} to tackle this challenge. However, unlike Graph Neural Networks (GNN), these models do not learn network structure information but instead receive it as input. As a result, the structure-based input needs to be regenerated whenever there is a modification in the network structure or the creation of a new firewall rule. This may entail retraining the entire DL model, introducing computational overhead.


Another limitation of existing approaches to network connectivity assessment for identifying zero-day vulnerabilities is overlooking mitigation techniques for the identified network threats \cite{ZD1, k-zero, patrol, ZDATT}. To effectively mitigate zero-day attacks, a coordinated defense is needed---one that includes both rapid and accurate risk assessment and a thorough prevention technology in the event a risk is identified.

\textbf{Challenges.} Assessing and mitigating the risks of zero-day attacks face three primary challenges stemming from the dynamic network connectivity structure being regulated by conventional firewall rules and NGFW policies: 
\begin{enumerate}[label=C-\arabic{enumi}., leftmargin=*]
    \item \textit{Adaptiveness}: Developing an automated and adaptive approach to identify high-risk connectivity paths that expose critical network assets, considering the dynamic nature of the network structure influenced by trends like remote users, bring-your-own devices, and cloud assets.
    \item \textit{Robustness}: Thoroughly identifying potential zero-day threats, considering that the vulnerabilities leading to zero-day attacks are inherently unknown to defenders, and attackers consistently discover new methods to exploit these vulnerabilities and network misconfigurations.  
    \item \textit{Proactiviness}: Effectively characterizing and ranking the risks associated with network connectivity paths that expose critical assets, enabling the proactive mitigation of those requiring prompt response without causing disruptions to network functionalities.
\end{enumerate}

\textbf{Our Work.} We introduce Pro-ZD, a framework designed for the autonomous identification of high-risk connections that could potentially be exploited by attackers to execute zero-day attacks. Specifically, Pro-ZD utilizes a GNN model to assess the network connectivity enabled by both traditional and NGFW policies. It evaluates the associated risks of these connections being exploited to compromise critical assets and incorporates proactive mitigation strategies for high-risk connections. Pro-ZD tackles the above challenges as follows:
\begin{enumerate}[label=(\arabic{enumi}), leftmargin=*]
\item 
To overcome the adaptiveness challenge, we introduce an innovative Graph Neural Network (GNN) method for identifying weighted paths. This dynamic approach characterizes paths from a vulnerable node (potential attacker entry point) to critical network assets. Leveraging the inductive property of GNNs, it efficiently generates node embeddings for previously unseen data. GNNs, incorporating network structural information as learnable features, exhibit self-adaptability to dynamic changes in the network structure.
\item 
Considering the inherent uncertainty of zero-day vulnerabilities, we address robustness challenges by shifting our threat identification focus to expose network connectivity misconfigurations within existing firewall rules. These misconfigurations may be potential entry points for attackers to reach vulnerable network assets. Our approach operates on the premise that preventing an attacker from establishing a connection to an asset hinders the exploitation of both known and unknown vulnerabilities. By mitigating high-risk connections to critical assets, we effectively counter zero-day threats without explicitly identifying the zero-day vulnerabilities.
\item 
To enhance efficiency, we automate the firewall rule risk analysis, evaluating their potential impact on exposing critical assets. This assessment considers network configuration, the criticality of assets, open port count, IP range of connection points, and path characteristics to high-criticality assets, adhering to governance rules for optimal network connectivity. We then autonomously mitigate high-risk connections by automatically adjusting firewall rules and Zero Trust (ZT) network policies (see Section~\ref{sec:ZT}) to disrupt critical paths without affecting overall network functionality.
\end{enumerate}

In this research, we tackle a significant limitation of existing GNNs that overlooks the capturing of positional information for nodes within the broader context of the graph structure~\cite{Spagan, distanceEncoding, SPGNN, PGNN}. Specifically, if two nodes share identical local neighborhood patterns but exist in different regions of the graph, their GNN representations become indistinguishable. To overcome this limitation, we introduce a novel GNN model for weighted shortest paths, named GraphWSP. 

The GraphWSP is an extension of our prior work, the Shortest Path Graph Neural Network (SPGNN) framework~\cite{SPGNN}. It utilizes the calculated SPGNN shortest path embeddings to compute {\it weighted} shortest paths to a predefined set of nodes representing critical network assets, where the weights are employed to characterize the risks associated with the identified paths. Unlike other approaches \cite{PGNN, distanceEncoding} that output a representation of the length of the paths, the SPGNN recovers the true shortest path distance from the node embeddings through a transferable GNN model ~\cite{SPGNN}. Nevertheless, the SPGNN does not account for edge features, therefore it is incapable of identifying the weighted shortest path length.

To facilitate the learning of path weight embeddings, we extend the SPGNN model by incorporating a stacked, innovative Graph Attention (GAT) network model. This ensemble is well-suited for analyzing intricate graph-structured data. The GAT network is specifically crafted to develop the ability to prioritize crucial neighbors, allowing it to highlight edges along the path to critical assets. Importantly, this prioritization is achieved without the need for computationally intensive matrix operations to compute the corresponding path weights.

\textbf{Contribution.} Our contributions are outlined as follows: 
\begin{itemize} [noitemsep, leftmargin=*, topsep=0pt]
\item We introduce a novel transferable GNN model for calculating weighted shortest paths, an extension of our prior work \cite{SPGNN} designed for shortest path computation. This approach can transfer previous learning of weighted shortest paths to previously unseen networks, mitigating challenges associated with the lack of labeled data.
\item We employ our proposed GNN model to formulate an innovative framework for assessing the risks associated with the automated generation of firewall rules. This framework identifies connections exposing critical assets susceptible to potential exploitation in zero-day attacks. Leveraging the inductive ability of GNNs, the model accommodates the dynamic nature of enterprise networks. 
\item The proposed framework includes proactive mitigation of potential zero-day attacks by automating the adjustment of network firewall rules and ZT policies. This process blocks high-risk connections exposing critical assets without disrupting network functionalities.
\end{itemize}


\textbf{Evaluation.} We compare the GNN weighted shortest path identification performance with the state-of-the-art GNN path identification model ``SPAGAN''~\cite{Spagan}. Results consistently demonstrate the superior performance of GraphWSP, achieving an average precision of over 85\% in the train-test setting and over 75\% in the transfer learning setting as compared to 63\% and 21\% for the SPAGAN, respectively.

Additionally, we evaluate the performance of Pro-ZD for identifying high-risk connections and associated firewall policies relying on data labeled by the network administrators of the organizations contributing to this work. The labeling pinpoints high-risk network connections requiring immediate intervention. Evaluation results attest to the robustness of Pro-ZD and its capacity to recover perturbations in the training dataset through topology-based label propagation (cf. \S~\ref{sec:evaluation}).


The rest of the paper is organized as follows: In Section~\ref{sec:rwork}, we survey the literature. In Section~\ref{sec:brwork}, we overview the ZT network architecture on which we base the connections risk assessment and mitigation. Section~\ref{sec:design} details the design of our Pro-ZD framework and Section~\ref{sec:experiments} elaborates on our experimental setup and datasets used for evaluating our proposed approach. We evaluate our model and present our results in Section~\ref{sec:evaluation}. Finally, Section~\ref{sec:conclusion} concludes our paper.

\section{Related Work}\label{sec:rwork}

In this section, we conduct a literature review 
about network connectivity structure-based risk assessment (see Section~\ref{sec:attackgr}), and GNN-based distance encoding and shortest path identification (refer to Section~\ref{sec:gnnsp}).
\subsection{Network Connectivity Risk Assessment}\label{sec:attackgr}

Prior studies have delved into the application of attack graphs and attack trees for cyber risk assessment ~\cite{byres, mcqueen, wang2013, AWAN2016}. These methodologies carefully examine the interdependencies between vulnerabilities and the network connectivity posture, guided by existing firewall rules and policies, to identify potential paths for attacks.


In \cite{AG_ZD, ZDATT}, the authors suggest that zero-day attacks often involve a mix of known and unknown exploits. Detecting zero-day attacks can then be accomplished by identifying known exploit signatures in network traffic. Hence, Attack graphs can be employed to predict zero-day attacks. Other approaches using attack graphs focus on evaluating risks linked to zero-day attacks. This involves predicting the number and locations of zero-day vulnerabilities, and analyzing network connectivity structure to assess the likelihood of attackers reaching and exploiting these predicted vulnerabilities \cite{ZD1, k-zero, patrol, ZDATT}.

This class of models, dependent on attack graphs, provides a structured perspective on potential threat scenarios within networks, leveraging the network connectivity structure. However, their effectiveness is tied to the network connectivity established by firewall rules and policies, and they are impeded by their inability to adapt dynamically to changes in the network connectivity structure. Any alteration to the network structure mandates the regeneration of the attack graph \cite{SPGNN}.

Numerous deep learning (DL) approaches have been proposed in the literature for various cybersecurity applications~\cite{DL_cyber, deeplearningattacks, deeplearningSecurity, DL_vul1, dl_vul2, dl_vul3}. Effective identification of potential network exploits necessitates a comprehensive understanding of the network structure and configurations. In DL-based approaches, unlike GNNs, the network structure information is not learned but is instead provided as input to the DL model. Consequently, the structure-based input must be regenerated whenever there is a change in the network structure, potentially requiring the retraining of the entire DL model.

GNNs have found recent applications in the cybersecurity domain, including vulnerabilities analysis \cite{Gnnvulnerability, Gnnvulnerability2}, botnet detection \cite{Gnnbots}, anomaly detection \cite{Gnnanomaly, GnnAnomaly2}, malware detection \cite{Gnnmalware}, and intrusion detection \cite{Gnnintrusion}. However, these methodologies exhibit limitations in their ability to conduct risk assessment and prioritize threats for mitigation.
\subsection{GNN Shortest Path Identification}\label{sec:gnnsp}

The objective of graph representation learning is to generate representation vectors that effectively encapsulate the structure and characteristics of graphs. This is particularly crucial as the expressive capability and accuracy of the learned embedding vectors significantly influence the performance of subsequent tasks like node classification and link prediction.

Nevertheless, existing GNN architectures exhibit limitations in capturing the positional information of a given node relative to other nodes in the graph~\cite{GNNWEAK1} (refer to Section~\ref{sec:expressive}). GNN updates the representation of each node iteratively by aggregating representations of its neighbors. Due to similar neighborhood structures shared by many nodes, the GNN may produce identical representations for them, even when these nodes are situated at different locations in the graph.

To address this limitation, recent works have attempted to introduce positional embeddings~\cite{distanceEncoding,PGNN}. However, this category of models primarily generates relative positional embeddings rather than actual values representing the shortest path length from a target node in the network.

Various methods for calculating the shortest path have been proposed in the literature. For instance, SPAGAN~\cite{Spagan} employs path-based attention in node-level aggregation to determine the shortest path between a central node and its higher-order neighbors. This approach allows for more effective information aggregation from distant neighbors into the central node. However, SPAGAN heavily depends on graph element features rather than positional embeddings, making its performance less optimal when limited features are available.

The SPGNN model~\cite{SPGNN} introduces the first GNN-based approach that is transferable and accurately calculates shortest paths using only distance information, without relying on other node or edge features. Nevertheless, it does not account for edge weights, rendering it unsuitable for weighted shortest path calculations. In this study, we extend the SPGNN model to enable the calculation of weighted shortest paths.

\section{Background}
\label{sec:brwork}
In this section, we provide an overview of the zero-trust architecture, along with related firewall policies and governance compliance, forming the basis for the risk assessment of network structure and the mitigation of identified high-risk connections (cf. \S~\ref{sec:ZT}, \ref{sec:gov}). 
Additionally, we explore the processing of graph data with GNNs and highlight the limitations of existing GNN architectures (cf. \S~\ref{sec:expressive}), which have motivated the development of our novel GNN-based model for weighted shortest path identification.

\subsection{Zero-Trust Architecture}
\label{sec:ZT}
Zero-trust (ZT) represents a comprehensive approach to securing corporate or enterprise resources and data, encompassing aspects such as identity, credentials, access management, hosting environments, and interconnecting infrastructure. The enactment of zero-trust (ZT) Architecture (ZTA) can take various forms within workflows. For example, micro-segmentation~\cite{ACSC} operationalizes ZTA by establishing secure zones in cloud and data-center environments, independently isolating and securing different application segments. This approach generates dynamic access network-layer control policies, which restrict network and application flows between micro-segments based on the characteristics and risk tolerance of the underlying network's assets.

Micro-segmentation is deployed through a distributed virtual firewall that governs access according to network-layer security policies for each micro-segment. By restricting access to only essential components, micro-segmentation plays a crucial role in preventing the proliferation of attacks within a network. The policies associated with ZT micro-segmentation (ZT policies) pertain to the network-layer policies enforced by the distributed firewalls for micro-segmentation, controlling the internal communications of the network. These policies are structured as follows: \code{< Source Micro-Segment IP Range > < Destination Micro-Segment IP Range > < Protocol > < Port Range >}.

\subsection{Governance and Compliance}
\label{sec:gov}
The optimal management of network communication policies relies on the visibility of underlying assets' characteristics and criticality within network micro-segments. For this purpose, a semantic-aware tier, referred to as "governance," is employed to ensure compliance of ZT policies with best practices for communication among network assets~\cite{noms2022}. The governance tier utilizes semantic tags (e.g., Database, Web Server, etc.) to conduct a risk-aware classification of micro-segments and their underlying assets, considering the criticality of the data stored, transported, or processed by these assets and the accessibility of the assets~\cite{assetValuation}.

In this study, we adopt eight criticality levels to classify network micro-segments based on the assigned workload. A workload refers to a collection of IT assets (servers, VMs, applications, data, or appliances) that collectively support a defined process (such as an application server, web server, or database). Network traffic among IT assets often provides a visualization of workloads. The criticality levels assigned to different network workloads are detailed in Table~\ref{tab:crit}. This table is generated in accordance with the investigation in~\cite{assetValuation}, supplemented by insights from the network security administrators of the two enterprises participating in this study.

A governance rule embodies the recommended practice governing which entities are permitted to communicate with different network assets. It relies on the assigned tags of micro-segments to evaluate the communications facilitated by the ZT policies within the network. The format of a governance rule is as follows: \code{< Source Tag > < Destination Tag > < Service Tag >}. It is essential to note that governance rules are crafted following optimal network communication practices and are tailored to each organization based on its network structure and business processes.

The Governance module evaluates the adherence of each ZT policy to its respective governance rule. Compliant connections refer to communications allowed by ZT policies that align with the stipulated governance rules. For example, if the ZT policy is \code{< Human-Resources Web Server IP Address > < Human-Resources Application Server IP Address > < TCP > < 443 >}, and it complies with the governance rule \code{< Web Server > < Application Server > < Secure Web >}, then all communications facilitated by this ZT policy are secure.

In a network context, connections labeled as "compliant" are typically deemed trustworthy in accordance with governance and access policies. Conversely, non-compliant connections may involve entities or applications whose identities are unknown to the governance module, leading to the absence of proper controls and access policies. As a result, non-compliant connections are considered less secure.


\begin{table}[h]
\centering

\tabcolsep=0.001cm
\begin{tabular}{|l|l|}
\hline
\textbf{Level} & \textbf{Description}\\
\hline
0 & UnTagged/unknown\\
1 & Untrusted and external/public e.g internet 0.0.0.0/0\\
2 & Trusted external e.g vendor\\
3 & Internet facing\\
4 & Untrusted and internal e.g users\\
5 & Internal \& connecting to untrusted internal e.g web servers\\
6 & Internal and connecting to data or non-critical data\\
7 & Critical data\\

\hline
\end{tabular}
\caption{\small Assets criticality levels and associated description.}
\label{tab:crit}
\vspace{-2mm}
\end{table}

In this study, our primary focus is on high-risk connections that can expose critical assets, particularly those involving non-compliant connections, which signify a relatively higher risk of exploitation. Here, critical assets are defined as network resources deemed valuable due to the sensitivity of the data they harbor, such as databases. Formally, $\mathit{V_{critical}= \{ v \;| \; v \in V \; \land \; c_{v} \;= \; 7 \}}$, where $\mathit{c_{v}}$ represents the criticality rating of node $\mathit{v}$ based on its governance tag.

\subsection{GNNs Expressive Power} \label{sec:expressive}

The objective of graph representation learning is to generate accurate graph representation vectors that capture the structure and features of graphs. Classical methods for learning low-dimensional graph representations, such as DeepWalk and Node2vec~\cite{deepwalk,Node2vec}, are inherently {\it transductive}. They make predictions on nodes in a single, fixed graph (e.g., using matrix-factorization-based objectives) and do not naturally generalize to unseen graph elements.

GNNs~\cite{semiKipf, kipf2} belong to the category of artificial neural networks designed for processing data represented as graphs. Unlike classical approaches, GNNs learn a function that generates embeddings by sampling and aggregating features from a node's local neighborhood, enabling the efficient generation of node embeddings for previously unseen data. This {\it inductive} approach is crucial for handling evolving graphs and networks that constantly encounter unseen nodes.

The effectiveness of neural networks lies in their strong expressive power, allowing them to approximate complex non-linear mappings from features to predictions. GNNs, in particular, learn to represent structure-aware embeddings for nodes in a graph by aggregating information from their k-hop neighboring nodes. However, GNNs have limitations in representing a node's position within the broader graph structure~\cite{distanceEncoding}. For example, nodes with topologically isomorphic local neighborhood structures and shared attributes but located in different parts of the graph will have identical embeddings.

The expressive power of GNNs is bounded by the 1-Weisfeiler-Lehman (WL) isomorphism test~\cite{GNNWEAK1}. In other words, GNNs exhibit limited expressive power, producing identical vector representations for subgraph structures that the 1-WL test cannot distinguish, even if they are substantially different~\cite{PGNN,distanceEncoding}.

\section{The Pro-ZD System}
\label{sec:design} 

\emph{Next}, we introduce our proposed framework designed to autonomously identify, characterize, and mitigate high-risk non-compliant network connections exposing critical assets. The Pro-ZD framework comprises three modules: (a) Network data pre-processing and feature extraction (cf. \S~\ref{sec:features}), (b) GNN-based model for computing weighted shortest paths to critical assets (cf. \S~\ref{sec:GraphWSP}), and (c) risk triage and proactive mitigation of high-risk connections (cf. \S~\ref{sec:criticality}). Further details on each module are provided in the subsequent subsections.

\subsection{Network Data Pre-processing}\label{sec:features}
We express a given network as a directed connectivity graph. Let $\mathit{C(V,E,S)}$ denote a labeled, directed connectivity graph, where $\mathit{V}$ corresponds to the set of graph vertices representing network assets (servers and cloud resources). The set of graph-directed edges $\mathit{E}$, where $\mathit{E \subseteq \{ (v,u,s) \;| \; (v,u) \in V^2 \land v \neq u \land s \in S\}}$, signifies communication between connected vertices utilizing the service identified through the edge label $\mathit{s \in S}$. $\mathit{S}$ denotes the set of network services defined by a protocol and port range.

\textbf{Features Extraction. }We derive the feature vectors that define the graph vertices (network assets) and edges (communication) from the packet headers of layers 3 and 4 network flow, the governance tags assigned under ZT, and the adherence of connections to governance rules. To evaluate the risk associated with network connections, we extract two types of features: (1) directly encoded from the network connectivity graph and (2) based on the network structure. 

\textit{a) Directly encoded features. }To incorporate the first category of features, we rely on the enterprise network-attributed connectivity graph. The directly encoded features include (1) the IP range of the edge (connection) source, (2) IP range of the edge destination, (3) number of open ports, (4) asset criticality of the edge source, (5) asset criticality of the edge destination, (6) the application criticality of the edge destination, (7) a boolean feature to indicate whether a node is directly connected to a critical asset and (8) edge compliance.

Features $FD_1$ and $FD_3$ indicate the likelihood and the number of ways a connection can be reached and exploited. Features $FD_4$ and $FD_8$ indicate the ease of connection accessibility. The lower the criticality of the source, the less the implemented constraints on its accessibility. Features $FD_2$, $FD_5$, $FD_6$, and $FD_7$ are an indication of the potential impact of having the connection exploited in terms of the number of potentially impacted assets and the importance of the data stored or processed on the impacted assets.

Feature $FD_6$ introduces a novel metric in this context, namely {\it Application Criticality (AC)}. Diverging from the asset’s criticality metric introduced in Section~\ref{sec:gov}, which categorizes assets based on their workload, the AC metric evaluates risk based on the application to which the asset belongs. For instance, a human-resources application database containing personally identifiable information receives a higher AC rating than a database associated with an inventory application.

Applications can be classified based on the scope of expected damages, if the application fails, as either, mission-critical, business-critical, or non-critical (operational and administrative)~\cite{app_criticality}. Mission-critical systems are essential for immediate operations, with even brief downtime causing disruptions and significant immediate and long-term consequences. Business-critical applications support long-term operations but do not necessarily result in an immediate disaster upon failure. Finally, organizations can continue normal operations for long periods without the non-critical application. Two different organizations might use the same application but it might only be critical to one. Hence, we rely on the security team of enterprises contributing to this study to assign the AC.

\textit{b)Network structure-based features. }The second feature category, rooted in the network structure, doesn't directly stem from the connectivity graph but is learned through a GNN model. Within this category, a solitary feature, denoted as $FS_1$, is introduced to discern the potential accessibility of a critical asset from the edge under evaluation. This feature gauges the likelihood of a zero-day attack utilizing this edge leading to the compromise of critical network assets. Hence, our model takes into account both immediate and cascading risks associated with the exploitation of non-compliant connections.

To express the exploitability of the identified path to critical assets, we evaluate the count of non-compliant edges on the path. As explained earlier in Section~\ref{sec:gov}, non-compliant edges inherently possess lower security levels and are more susceptible to exploitation. Consequently, $FS_1$ focuses on assessing the weighted shortest path to critical assets, where weights indicate the compliance status of the path's edges. 

Given that the overall path weight serves as an indicator of path length, a relatively low weight signifies the ease of exploiting the path to reach a critical asset. In alignment with this, a compliant edge is assigned a weight relatively higher than that of a non-compliant edge. In this work, we adhere to the criteria outlined by the network administrators of the participating enterprises, defining a path as easily exploitable if its length is less than four edges and it contains at least two non-compliant edges leading to a critical asset.

\textbf{Weighted adjacency matrix: }For the computation of weighted shortest paths, our primary focus is on edges' compliance. Other edge features, such as service port and protocol, are excluded from $FS_1$ as they are part of the directly encoded features. To ensure a representative adjacency matrix that reflects edge weights, we opt to eliminate redundant edges. In essence, if there are multiple edges with the same compliance type connecting two nodes, we retain only one. Consequently, considering there are two potential values for compliance (either compliant or non-compliant), the maximum number of edges between two nodes is two.

In calculating the weighted adjacency matrix $\mathcal{A}$ for the connectivity graph $\mathit{C}$, when there is more than one edge connecting two nodes, the value in the adjacency matrix is the sum of the weights of the two edges. Given that there are only three possibilities for edges connecting two nodes---either one non-compliant edge, one compliant edge, or a combination of compliant and non-compliant edges---the weights in the adjacency matrix remain representative.

\subsection{GraphWSP: Weighted Shortest Paths GNN}\label{sec:GraphWSP}

The GraphWSP model generates weighted shortest path embeddings from a specific node in the network to a predefined set of anchor nodes. The anchor nodes play a crucial role in determining the relative position of a node within the graph and can be selected either randomly with a specific distribution or based on node features. In our context, the anchor nodes represent the set of critical network assets. 

The identification of weighted shortest paths follows a three-step approach. Firstly, we calculate the shortest path length to a predefined set of anchor nodes using the SPGNN model~\cite{SPGNN}. Secondly, we stack a GAT model to generate path weight embeddings based on the network-weighted adjacency matrix, where the weights signify the connection compliance with the governance rules (see Section~\ref{sec:gov}). Thirdly, these weighted shortest path embeddings are integrated into the downstream classification task, where we classify the criticality of paths to discern those necessitating immediate attention.

For efficient graph representation learning, utilizing node features is more effective than edge features, given the richer information content in nodes and their relatively smaller number compared to edges. We start by predicting weighted shortest paths as additional node features and then associate the calculated distance with all incident edges of the node. Let $\mathit{v}$ be a node, $\mathit{FS_1(v)}$ the learned representation of its weighted shortest path, and $\mathit{y_e}$ the feature vector for edge $\mathit{e}$. Node features are then assigned to incident edges as follows:
\begin{equation}\label{eq:edgeFeature}
 \{\forall u \in V \; \land \; \exists \;e_{u,v} \in E\;, \;y_{e_{u,v}} = FS_1(v)\} 
\end{equation}

\textbf{(1) Shortest Path Length: }
The SPGNN model is designed to predict the shortest path length to a predefined set of anchor nodes, specifically highly-critical assets, denoted as $\mathit{V_{critical}}$, within the network. The goal of the SPGNN is to learn a mapping $\mathit{V \times V_{critical}^{k}} \mapsto  R^{+}$ to predict the actual minimum shortest path distances from each node in the network $\mathit{u \in V}$ to $\mathit{V_{critical}}$ where $\mathit{k = |V_{critical}|}$. Since the model is designed to assess the risk associated with critical paths, we address the worst-case scenario by evaluating the minimum shortest path length from the vulnerable node to a highly-critical asset. Consequently, the loss is calculated based solely on the minimum value within the shortest paths distance vector.

The objective for learning the actual shortest path length is formulated as follows~\cite{SPGNN}, where $\mathcal{L}$ is the mean squared error (MSE) loss function to be minimized:
\begin{equation} \label{eq:loss}
\begin{aligned}
&\min_\phi \sum_{\forall u \in V} \mathcal{L}\left( \min_{i \in\{1\dots k\}}\hat{d}_\phi\left(u, v_{i}\right)-\min_{i \in\{1\dots k\}} d_{y}\left(u, v_{i}\right)\right)\\
&\min_\phi \sum_{\forall u \in V} \mathcal{L}\left(\min \left(\hat{d}_\phi\left(u, V_{critical}\right)\right)-\min \left(d_y(u, V_{critical}\right)\right)
\end{aligned}
\end{equation}
Here $\mathit{\hat{d}_\phi\left(u, V_{critical}\right)}$ is the vector of learned approximation of the shortest path distance from a node $\mathit{u}$ to every critical asset $\mathit{v} \in V_{critical}$. $\mathit{d_{y}}$ is the observed shortest path distance.




In this approach, the message-passing function relies solely on positional information to compute absolute distances to the anchor sets, without considering node features. To calculate position-based embeddings, the 1-hop distance $\mathit{d_{sp}^1}$ can be directly derived from the adjacency matrix. During the training process, the shortest path distances $\mathit{d_{sp}^q(u,v)}$ between a node $u$ and an anchor node $v$ are calculated as follows~\cite{SPGNN}, where $\mathit{d_{sp}(u,v)}$ is the shortest path distance:
\begin{equation}\label{eq:messagePassing}
\begin{aligned}
d_{sp}^q(u,v) \mapsto   \begin{cases} 
      d_{sp}(u,v), & if\; d_{sp}(u,v) < q \\
      \infty & otherwise. 
   \end{cases}
\end{aligned}
\end{equation}

The model maps close nodes (in position) in the network to similar embedding. Accordingly, the distance is further mapped to a range in $\mathit{( 0,1)}$ as follows~\cite{SPGNN}:
\begin{equation}\label{eq:message}
s(u,v) = \frac{1}{d_{sp}^q(u,v)+1}
\end{equation}
Hence, the message-passing process can be defined as: 
\begin{equation}\label{eq:positionEmbedding}
    h_u=\phi(x_u \oplus_{(v \in \aleph_v)}\psi(u,v))
\end{equation}
Here, $\mathit{h_u}$ represents the node embedding of vertex $\mathit{u}$, $\mathit{x_u}$ is the input feature vector of node $\mathit{u}$ inferred from the adjacency matrix, and $\mathit{\oplus}$ denotes the aggregation function. In our approach, the mean aggregation function yields the best performance. $\mathit{\psi}$ represents the message function, computed as described in Equation~\ref{eq:message}. Finally, $\mathit{\phi}$ is the update function used to obtain the final representation of node $\mathit{u}$.

To recover the true path length from the learned node embedding, the model introduces four steps to the learning process after generating the node embeddings through message passing: Firstly, for every node $\mathit{u \in V}$, we calculate a vector of absolute distances (AD) $\mathit{T_u}$ between the learned embedding of $\mathit{u}$ denoted as $\mathit{h_u}$ and the embedding of every critical asset $\mathit{v_i \in V_{critical}}$, denoted as $\mathit{h_{v_i}}$. $\mathit{h_u}$ and $\mathit{h_{v_i}}$ are calculated as described in Equation~\ref{eq:positionEmbedding}.The AD vector is calculated as follows, where $\mathit{k=|V_{critical}|}$ is the embedding space dimension ~\cite{SPGNN}:
\begin{equation}
\begin{aligned}
AD (u,v) = \sum_{n=1}^{k}|h_u^{n} - h_v^{n}|\\
T_u = \forall_{v_i \in V_{critical}} AD(u,v_i)
\end{aligned}
\end{equation}
$\mathit{T_u}$ is then used in Equation~\ref{eq:loss} to calculate the loss where $\mathit{\hat{d}\left(u, V_{critical}\right)=T_u}$.

Secondly, given that the downstream task focuses on discerning the risk posed by potential paths, if a node $\mathit{u \in V}$ has (shortest) paths to multiple critical assets, we consider the worst-case scenario by determining the minimum length of the shortest paths $\mathit{T_u}$. Hence, the model attributes the minimum calculated AD to the node $\mathit{u}$. Thirdly, the model approximates the assigned AD to an integer value to represent the predicted shortest path distance. Lastly, it attributes the approximated shortest path value to the incident edge features.

\textbf{(2) Weighted Shortest Path: }To acquire path weight embeddings, we introduce an innovative GNN model that expands upon the SPGNN approach by incorporating a stacked novel graph attentional network (GAT) model for computing weighted shortest path embeddings. In the GAT model, the weight assigned for neighborhood aggregation is not fixed; instead, each edge in the adjacency matrix is assigned an attention coefficient, signifying the influence of one node on another. The network is designed to learn to prioritize important neighbors \cite{GAT}, enabling it to emphasize edges along the path to critical assets without necessitating computationally intensive matrix operations (e.g., inversion).


The attention function can be conceptualized as a mapping of a query and a set of key-value pairs to an output. As all the components---query, keys, values, and output---are vector-based, the output is derived through a weighted sum of the values. The weight for each value is decided by a compatibility function between the query and its associated key\cite{multihead}. 

To enhance the stability of the self-attention learning process, we incorporate multi-head attention. This involves projecting the queries, keys, and values multiple times through distinct learned linear projections. The attention function is then independently applied to each of these projected versions in parallel. The attention outputs are concatenated and linearly transformed into the expected dimension.

Let $\mathit{C(V,E,S)}$ represent the network connectivity graph, where $\mathit{V}$ stands for network assets and $\mathit{E}$ is the set of graph-directed edges indicating communication between connected vertices. Let $\mathcal{A}$ denotes the weighted adjacency matrix for the network connectivity graph $\mathit{C}$ where $\mathit{\mathcal{A}_{ij} \in \{0,\mathnormal{w}\}}$ denotes whether there is an edge going from node $\mathit{i}$ to node $\mathit{j}$ and $\mathnormal{w}$ is the edge weight reflecting the compliance.

Let $\mathit{H=\{\overrightarrow{h_1},\overrightarrow{h_2},...,\overrightarrow{h_N}\}}$ be the set of node features, with $\mathit{\overrightarrow{h_i} \in \mathbb{R}^{F_H}}$. Here, $\mathit{N =|V|}$ and $\mathit{F_H}$ is the number of node features. The node features encompass both directly encoded features $FD_4$, $FD_7$, and $FD_8$ and a structure-based feature corresponding to the output of the SPGNN denoted as $\mathit{SP \in \overrightarrow{h_i}}$. Let $\mathit{Y=\{\overrightarrow{y_1},\overrightarrow{y_2},...,\overrightarrow{y_M}\}}$ be the set of edge features, where $\mathit{\overrightarrow{y_p} \in \mathbb{R}^{F_Y}}$, $\mathit{M =|E|}$, and $\mathit{F_Y}$ is the number of edge features. In this study, we specifically consider one edge feature, which corresponds to edge compliance where for an edge ${e_{ij}}$, $\mathit{\mathcal{A}_{ij}}$ $\mathit{\in \overrightarrow{y_{e_{ij}}}}$. 


At each GAT network layer, a learnable linear transformation parameterized by a weight matrix $\mathit{W}$ is required to transform the input layer features $\mathit{H}$ into higher-level features $\mathit{H'}$, potentially with different cardinality $\mathit{\overrightarrow{h'_i} \in \mathbb{R}^{F'_H}}$:
\begin{equation}
\overrightarrow{h'}_{i}^{l}= W^{(k)l} \quad \overrightarrow{h}_{i}^{(l-1)}
\end{equation}
Where $\mathit{W^{(k)l}}$ is defined for each layer $\mathit{l}$ and attention head $\mathit{k}$. $\mathit{\overrightarrow{h}_{i}^{0}}$ represents the node $\mathit{i}$ feature vector of the model input. 

An attention mechanism is then performed on each node $\mathit{i}$, generating the attention weights of its neighbors that indicate the importance of each node $\mathit{j}$’s features to node $\mathit{i}$. In our approach, the attention mechanism $\mathit{\alpha^{(k)l}_{ij}}$ for a layer $\mathit{l}$ and attention head $\mathit{k}$ is a single-layer feed-forward neural network, parameterized by a weight vector $\mathit{\overrightarrow{a} \in \mathbb{R}^{2F'_H}}$.

During the process, node features are concatenated and the $\mathit{LeakyReLU}$ activation function is applied. The coefficients computed by the attention mechanism include not only the features of the two nodes but also the weight of the edge connecting them where $\overrightarrow{y_{e_{ij}}}$ is the edge $e_{ij}$ feature vector that includes the weight of the edge in the adjacency matrix. While this approach enables a node $\mathit{i}$ to attend to every other node in the graph, we perform masked attention to limit the computation of weights to the set of first-order neighbors $\mathcal{N}_i$ of node $\mathit{i}$. The attention mechanism can be expressed as:
\begin{equation}
\alpha^{(k)l}_{ij} = \frac{exp(\overrightarrow{y_{e_{ij}}}\sigma(\overrightarrow{a}^T [\overrightarrow{h'}_{i}^{l} || \overrightarrow{h'}_{j}^{l}]))  }{\sum\limits_{g\in\mathcal{N}_i} exp(\overrightarrow{y_{e_{ij}}}\sigma( \overrightarrow{a}^T [\overrightarrow{h'}_{i}^{l} || \overrightarrow{h'}_{g}^{l}]))}
\end{equation}



A forward process, for updating node features based on the computed attention, is subsequently executed. It calculates a linear combination of features corresponding to the determined attentions, serving as the final output features $\overrightarrow{h}_{i}^{l}$ for each node, post-application of a non-linear transformation $\Phi$.

We extend our mechanism to incorporate multi-head attention, where $D$ represents the number of attention heads. The $D$ independent attention features are then consolidated using $\Theta$, denoting the mean operation, yielding the following output feature representation:
\begin{equation}
\overrightarrow{h}_{i}^{l}=\Phi\{\Theta_{k=1}^{D} \{\sum_{\forall j \in \mathcal{N}_i} \alpha^{(k)l}_{ij} \quad \overrightarrow{h'}_{j}^{l-1}\}   \}
\end{equation}

The devised computation of weighted shortest path embeddings primarily serves downstream classification tasks. In this context, our objective is to categorize edges based on their adherence to the conditions stipulated by the structure-based feature outlined in Section \label{sec:features}. Initially, $FS_1$ is introduced to assess the potential accessibility of a critical asset from the edge under examination given the path length and the number of non-compliant edges on the paths. Thus, we configure the output layer with two neurons and apply a Softmax activation function. Moreover, we set a threshold to define the conditions for categorizing a path as easily exploitable. Our experiments demonstrate that, in this context, this configuration surpasses the performance of a single output class with a Sigmoid activation. 


Let $\mathit{Y}$ be the set of output labels representing the potential accessibility of a critical asset from the edge under examination. The goal of the GAT is to learn a mapping $\mathit{V \mapsto Y}$. Let $\mathit{f_{\theta} : V \mapsto Y }$ be the GAT network that maps the nodes to the set of labels $\mathit{Y}$ where $\mathit{\theta}$ denotes the parameters of $\mathit{f_{\theta}} $. The objective function for the nodes classification task can be formulated as minimizing the cross entropy loss function $\mathcal{L}$ as follows:
\begin{equation}
\min_\theta \sum_{\forall u \in V} \mathcal{L}\left( f_{\theta} \left( C(V,E,X,Y)\right),y_i|v_i \right)
\end{equation}

\begin{figure}[h]
    \centering
     
    \includegraphics[width=1\columnwidth]{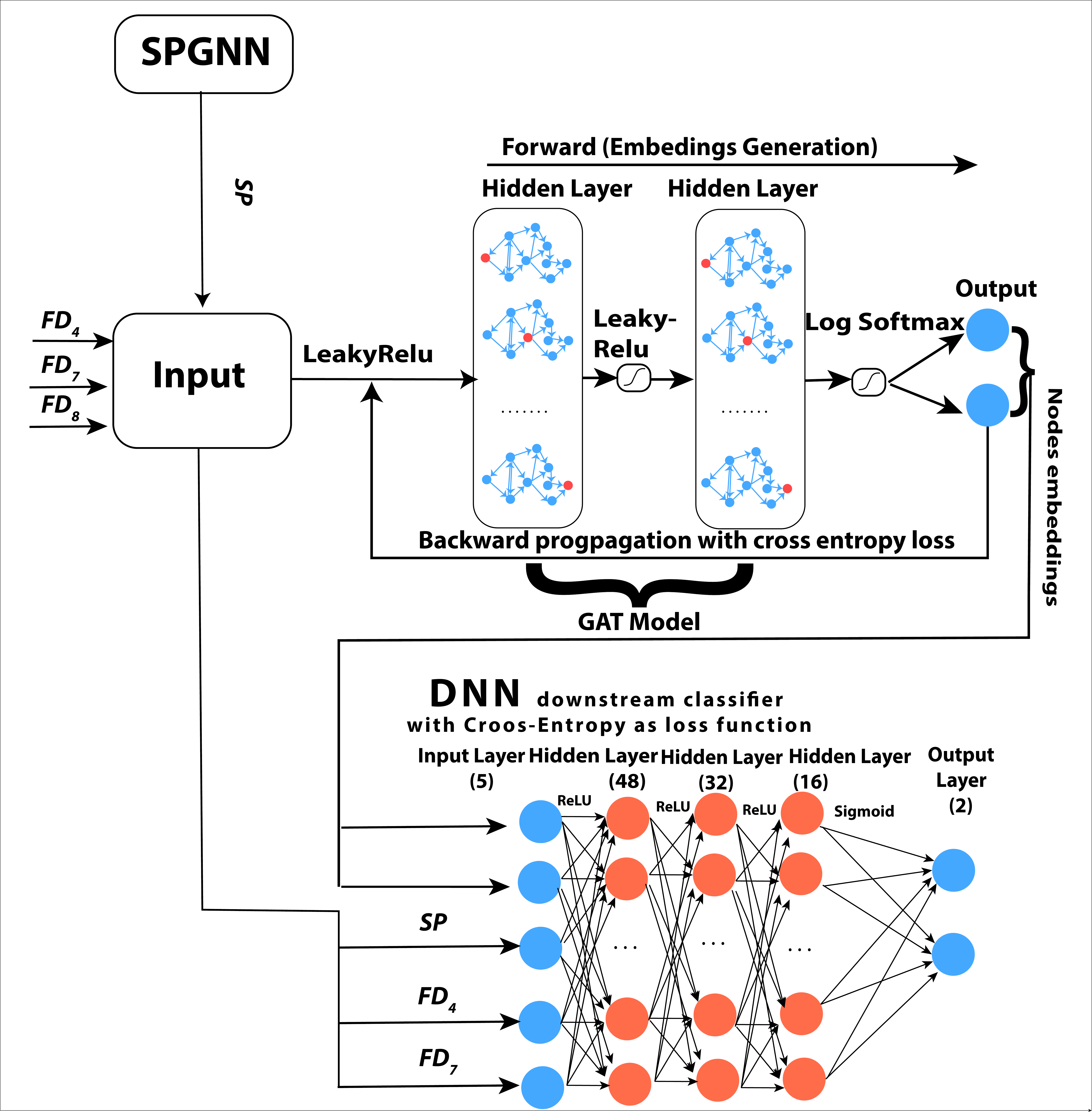}
   \caption{\small GraphWSP Architecture. In this diagram, the upper GNN depicts the GAT model employed for acquiring the weighted shortest path embedding. It receives four input features (asset criticality $FD_4$, link to critical asset $FD_7$, edges compliance $FD_8$, and shortest path length computed by the SPGNN $SP$). The lower section illustrates the DNN responsible for the downstream classification task. The output indicates whether the structure feature is activated for the edge under evaluation. The DNN takes as input the GAT output, asset criticality $FD_4$, link to critical asset $FD_7$, edges compliance $FD_8$, and the shortest path length $SP$ computed by the SPGNN.}
    \label{fig:self_supervised}
    \vspace{-2mm}
\end{figure}

\textbf{(3) Downstream Classification: }The output of the GAT is a $c$ dimensional array, where $c$ is the number of potential node classes. Given that this model mostly relies on node positional information and assumes limited node features, we observe that the GAT model output classification threshold is not consistent and requires learning. Accordingly, we stack a DNN classifier to the output of the GAT. 

The input of the DNN classifier comprises the GAT output, recording the node classes based on weighted shortest path embedding, in addition to the model's original input features to capture the nodes’ fundamental properties. The stacking of a DNN classifier has proven to be effective in improving the overall accuracy of the nodes classification and extending the model transferability.

The goal of the DNN is to learn a mapping from each node $\mathit{u \in V}$ to a label $\mathit{l_u \in L}$. Accordingly, the objective function can be represented as follows:
\begin{equation}
 \min_\phi \sum_{\forall u \in V} \mathcal{L}\left( g_\phi \left( \overrightarrow{\lambda_{u}}\right) ,l_u \right)   
\end{equation}
where $\mathit{g_\phi}$ is a function that maps the node features vector $\mathit{\overrightarrow{\lambda_{u}}}$, including the GAT output, to a label $\mathit{l_u \in L}$. $\mathit{\phi}$ denotes the parameters of $\mathit{g_\phi}$ and $\mathcal{L}$ is the cross entropy loss function.


Subsequently, for every node $\mathit{v \in V}$, the acquired labels $\mathit{l_v \in L}$ are assigned to its incident edges based on Equation~\ref{eq:edgeFeature}. This label serves as the structure-based feature $FS_{1}$ and is employed for the risk triage classification task.



\subsection{Risk Triage and Mitigation}\label{sec:criticality}

\begin{figure}[h]
    \centering
     
    \includegraphics[width=0.6\columnwidth ,height=0.45\columnwidth]{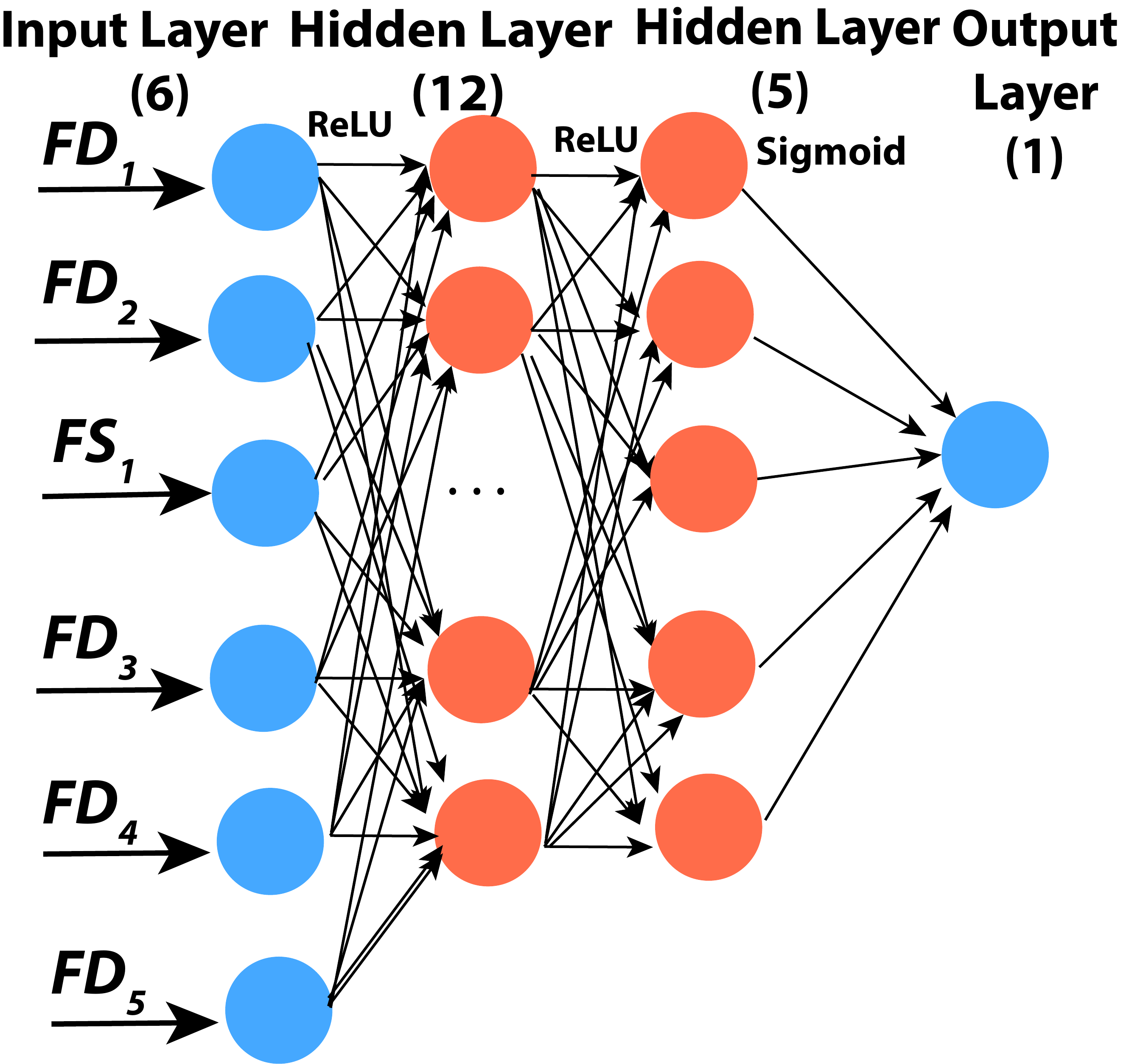}
   \caption{\small Edge Risk Assessment Model Architecture. The model takes as input the predicted weighted shortest path embedding $FS_{1}$ from GraphWSP model and 5 directly encoded features $FD_{i}$ }
    \label{fig:DNNfinal}
    \vspace{-2mm}
\end{figure}


We design a module to automate the evaluation of risk associated with potential exploitation of non-compliant connections exhibiting a detected easily exploitable path to critical assets. In this context, risk is characterized by the propensity and ease of an edge to be compromised, leading to the exposure of critical assets to zero-day attacks. Initially, we identify critical non-compliant edges necessitating immediate intervention based on the input features of the model. Subsequently, we autonomously pinpoint the policy facilitating the connection and take proactive measures to adjust the policies, thereby blocking the identified high-risk connection.

We devise a DNN edge classifier to evaluate the criticality of non-compliant connections, leveraging the directly encoded features $FD_1$ to $FD_5$ detailed in Section \ref{sec:features}, and the structure-based feature $FS_1$ acquired through the GraphWSP model outlined in Section\ref{sec:GraphWSP}. We define two output classes (labels), guiding subsequent mitigation actions: (1) Non-compliant critical, and (2) Safe.


Let $f_\psi: E \mapsto \mathit{Y}$ be a function that maps the set of edges $\mathit{E}$ to the set of labels $\mathit{Y}$ representing the two edge classes where $\mathit{\psi}$ denotes the parameters of $\mathit{f_\psi}$. Let $feat_{e}$ be the input feature vector of the edge $\mathit{e}$. To optimize the edge's classification task, we express the objective function as the minimization of the cross-entropy loss function $\mathcal{L}_{d}$, as follows: 
\begin{equation} \label{eq:nn_Loss}
\min_{\psi}\sum_{\forall e in E } (f_{\psi}(feat_{e}),y)
\end{equation}


\subsection{Data Augmentation}

We leverage data augmentation to enhance the generalizability of GraphWSP. This approach not only addresses potential imbalances in the data but also strengthens the model's capacity to comprehend the skewed relationships among graph nodes and edges. To implement this, we segment the network connectivity graph into a series of overlapping sub-graphs, each characterized by a predefined radius $\mathit{R}$. Subsequently, we supply the features of these sub-graphs to both the GAT and DNN classification models during the training phase.

The retrieval of sub-graphs involves iterating over all connections of each node within the specified radius $\mathit{R}$. Consequently, the model assimilates local structural information from subgraphs, considering pairwise details from multiple interconnected nodes and edges. Each sub-graph encapsulates a local perspective around a specific node, enriching the model's ability to capture local interactions effectively.



\section{Experiments}\label{sec:experiments}

\subsection{Datasets}\label{sec:data}

We evaluate the proposed model using two distinct classes of datasets: (1) Real-life datasets obtained from medium-sized enterprise networks, denoted as $RTD_1$ and $RTD_2$. (2) Synthetic datasets, labeled as $STD_1$ and $STD_2$, designed to mimic the characteristics of medium-sized enterprise networks. Table~\ref{tab:Data_stat} provides a comprehensive overview of the dataset details used for evaluating our model's performance.

For the real-world datasets ($RTD_1$ and $RTD_2$), we acquire data from two distinct medium-sized enterprises—an esteemed law firm and a university. The Nessus scan output serves as a valuable resource for discerning the configurations and properties of network assets. Additionally, we leverage firewall rules and ZT policies provided by the enterprises to characterize the communication patterns among assets.

To simulate medium-sized enterprise network scenarios, we generate two synthetic datasets ($STD_1$ and $STD_2$). These datasets are crafted to encompass all conceivable combinations of feature values pertinent to the assessment of network edge criticality. Each connection in the synthetic datasets is labeled with a binary tag, indicating whether it qualifies as a high-risk connection warranting automatic blocking for enhanced cybersecurity. The label assignment process is semi-automated, aligning with the procedures defined by the network administrators of the enterprises contributing to this research.

\begin{table}[h]
\centering
\scalebox{0.9}{
\begin{tabular}{l||r|r|r|r|r}
\hline
\toprule 
\textbf{Dataset} & \textbf{Nodes}& \textbf{Edges} & \makecell{\textbf{Critical} \\ \textbf{Assets}} & \makecell{\textbf{Compliant} \\ \textbf{Edges}} & \makecell{\textbf{Non-compliant} \\ \textbf{Edges}}\\
\hline
$SDT_1$ & 864& 5,018 & 284 &2,002&3,016\\
$SDT_2$ & 865& 5,023 & 284 &2,002&3,021\\
$RTD_1$ & 221& 1.914 & 21 &882&1,032\\
$RTD_2$ & 370& 21,802& 70 &10901&10901\\
\hline\bottomrule       
\end{tabular}
}
    \caption{\small Dataset features and statistics.} 
    \label{tab:Data_stat}
        \vspace{-4mm}
\end{table}

\subsection{Baseline Models}
 

We perform a comparative analysis between our proposed model, GraphWSP, and the state-of-the-art baseline SPAGAN~\cite{Spagan}, specifically focusing on the identification of weighted shortest paths. SPAGAN employs path-based attention, explicitly considering the influence of a sequence of nodes to determine the minimum cost or shortest path between the central node and its higher-order neighbors.

Given that the Pro-ZD establishes a novel approach for evaluating network connectivity and identifying potential risks associated with zero-day attacks, there are no existing baseline models suitable for benchmarking. To assess the performance of Pro-ZD in connectivity risk assessment and the identification of potential zero-day attack threats, we rely on accuracy and precision metrics, comparing the results against semi-automatically labeled datasets. In essence, the Pro-ZD model labels connections as either critical or safe based on the propensity of being exploited to compromise critical assets through a zero-day attack. The predicted labels are evaluated against manually labeled edges of the four datasets. 
\subsection{Experiment Setup}
In this study, we assess the performance of the proposed Pro-ZD framework and its underlying GraphWSP for weighted shortest path calculation across three distinct settings: 

{\bf Experiment 1 -- Evaluation of Weighted Shortest Paths Identification in a Semi-Supervised Setting. } This experiment focuses on gauging the capability of GraphWSP to identify weighted shortest paths in a semi-supervised context. The evaluation involves comparing prediction accuracy with the baseline model SPAGAN. To identify the minimum ratio of labeled data essential for satisfactory performance, we utilize train and test split masks with distribution shifts across all datasets detailed in Section~\ref{sec:data}. In this transductive learning setting, the model undergoes training and testing on a fixed graph with a predetermined node order.


{\bf Experiment 2 -- Assessment and Validation of Learning Transferability.} This experiment specifically addresses the assessment of learning transferability within the proposed GraphWSP for weighted shortest path identification. The model is tested for transferability by training it on one dataset and evaluating it on a distinct unlabeled dataset. We demonstrate the inductive learning performance of the model on node classification tasks, showcasing its ability to transfer positional information to new, unseen graphs.


{\bf Experiment 3 -- Assessment and Validation of Attack Paths Identification.} In this experiment, we aim to evaluate the end-to-end performance of the Pro-ZD framework in identifying high-risk network connections that expose critical assets. The evaluation involves testing the model output against labeled synthetic network datasets and real-world datasets sourced from enterprises contributing to this research. 

\section{Results \& Evaluation}\label{sec:evaluation}

The evaluation encompasses three distinct aspects: (1) assessing the performance of the GraphWSP in weighted shortest path calculation within a semi-supervised transductive learning setting (cf. \S~\ref{sec:evsp}), (2) evaluating the performance in a transfer-learning inductive learning setting (cf. \S~\ref{sec:transfer}), and (3) gauging the accuracy of identifying high-risk connections that expose critical assets (cf. \S~\ref{sec:attack}).

\subsection{Evaluation of GraphWSP in Transductive Setting}\label{sec:evsp}
To provide quantitative insights into the model's performance, we present the mean accuracy, F1 score, and receiver operating characteristic (ROC) curve derived from 100 runs and 100 epochs with diverse train/test masks.

\textbf{Accuracy evaluation. }The performance GraphWSP in comparison to the baseline SPAGAN is summarized in Table~\ref{tab:evaluation_feat3}. In general, GraphWSP surpasses the baseline model, effectively capturing the intricate relationships within the datasets, particularly evident in the assessment of real-world datasets. Moreover, the transformation of path length prediction into a classification task, utilizing one-hot encoding for output representation, allows the model to grasp the ordinal relationships between different lengths, resulting in enhanced performance. Both architectures exhibit a decline in performance when evaluated on real-world datasets, attributed to the challenging manual process of generating ground-truth labels, introducing inconsistencies to the unobserved latent labels. 

\begin{table}[h]
\tabcolsep=0.13cm
\scalebox{.78}{
\begin{tabular}{lcccc|cccc}
\hline
\toprule
         & \multicolumn{4}{l}{\bf SPAGAN} & \multicolumn{4}{l}{\bf GraphWSP} \\
         \cline{2-4}\cline{4-9}
{\bf Metrics}   & $STD_1$          & $STD_2$  & $RTD_1$ & $RTD_2$         & $STD_1$         & $STD_2$   & $RTD_1$ & $RTD_2$
\\ \hline
Accuracy & $87.860\%$             & $90.75\%$ & $82.22\%$& $82.43\%$           & $87.50\%$             & $87.90\%$ & $83.71\%$    &  $83.73\%$     \\
\hline
F1       & $82.35\%$             & $85.18\%$ & $0.0\%$ & $48.00\%$           & $83.65\%$             & $84.36\%$          & $77.77\%$  & $51.62\%$\\\hline
RoC\_Acc. & $86.40\%$             & $85.18\%$      & $50.00\%$   & $65.77\%$    & $88.71\%$             & $86.12\%$ & $54.91\%$  & $68.21\%$  \\ \hline\bottomrule       
\end{tabular}
}
    \caption{\small Evaluation overview the weighted shortest path prediction of GraphWSP versus SPAGAN for the four datasets.}
    \label{tab:evaluation_feat3}
    \vspace{-2mm}
\end{table}


\textbf{Configuration assessment. }For a more intuitive grasp of the proposed model's robust and promising performance, we delve into the exploration of various configuration schemes during the training stage. Specifically, we analyze the impact of varying train-test masks and the radius $\mathit{R}$, representing the size of subgraphs employed for data augmentation.

Upon testing the model with different subgraph sizes $\mathit{\in [1,6]}$, we observe that the performance gain achieved by increasing the radius beyond three hops is marginal across all datasets. This phenomenon can be attributed to the datasets' nature, where nodes beyond four hops from a critical asset are considered low risk. Consequently, for efficiency considerations, we set $\mathit{R = 3}$.

Figure~\ref{fig:F1_score_WSPGNN} illustrates the model's performance under different train/test masks, assuming $\mathit{R = 3}$. The graph reveals two peaks with nearly identical F1 scores at 40\% and 80\% training ratios for the $RTD_2$ dataset. No significant improvement is observed for the remaining datasets beyond a training ratio of 40\%. Hence, the results presented in this section are grounded in the configuration values of $\mathit{R = 3}$ and a training ratio of 40\%, unless explicitly stated otherwise.

\begin{figure}[h]
    \centering
     
    \includegraphics[width=1\columnwidth, height=0.5\columnwidth]{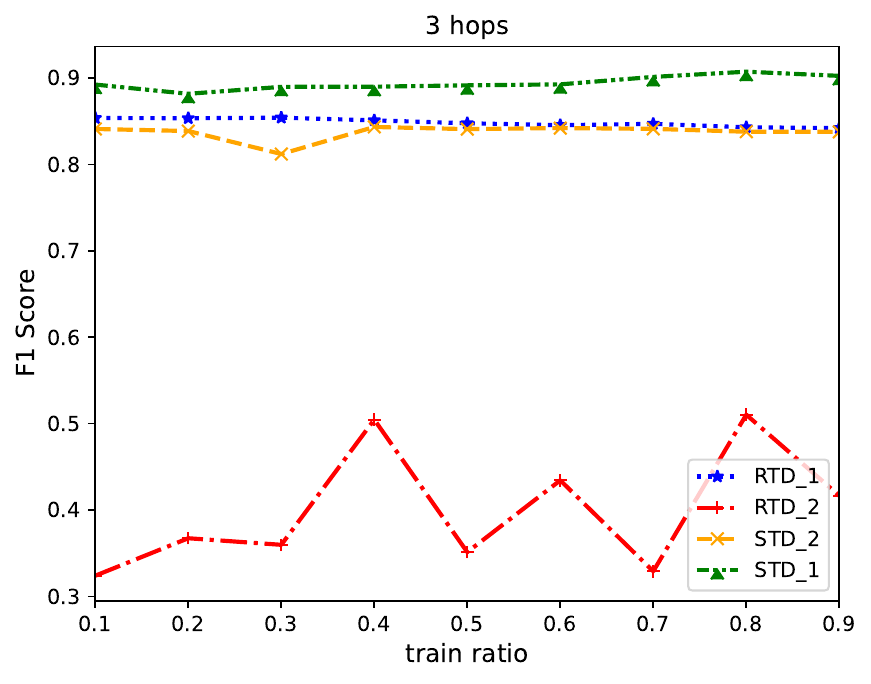}
    \caption{\small Train/test cut off for the W-SPGNN based on the F1 score for 4 different datasets}
   
    \label{fig:F1_score_WSPGNN}
    \vspace{-2mm}
\end{figure}

\textbf{Model convergence. }Figure~\ref{fig:loss_WPSGNN} illustrates the evolution of cross-entropy loss throughout the training iterations of WSPGN. The gradient exhibits significant magnitude, consistently moving in the direction of the steepest descent, effectively minimizing the objective function. This stability in the learning process contributes to the model's accuracy, demonstrating its robust performance across diverse dataset characteristics. Moreover, the smoothness of the objective function indicates that the model is not under-fitting.

\begin{figure}[h]
    \centering
     
    \includegraphics[width=1\columnwidth, height=0.5\columnwidth,]{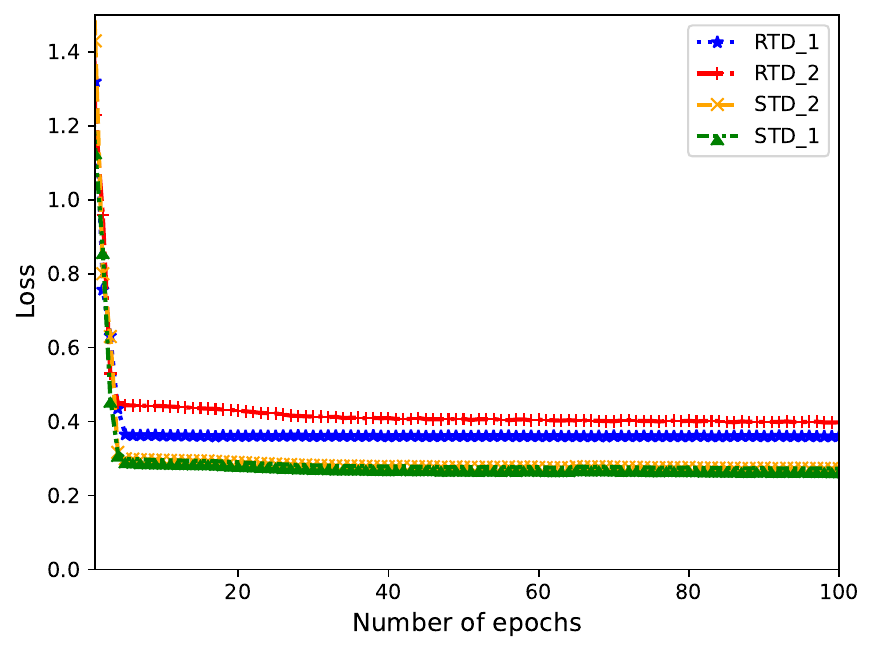}
    \caption{\small GraphWSP Cross Entropy loss convergence.}
   
    \label{fig:loss_WPSGNN}
    \vspace{-4mm}
\end{figure}

\subsection{Evaluation of GraphWSP in Inductive Setting}\label{sec:transfer}

Distinct datasets are employed for the pre-training and testing phases. The objective of pre-training is to transfer knowledge acquired from labeled datasets to facilitate downstream tasks with unlabeled datasets. To assess the robustness of this transferability, we pre-train the model using various synthetic and real-world datasets. Notably, the size and complexity of the dataset used for pre-training significantly influence the transferability performance. As indicated in Tables \ref{tab:WSPGNN evaluation} and \ref{tab:evaluation Transfer SPAGAN}, the superior performance of training with $STD_1$ is evident.

A comparative analysis of GraphWSP and SPAGAN is conducted. The F1 score values in Tables \ref{tab:GraphWSP evaluation} and \ref{tab:evaluation Transfer SPAGAN} reveal the better performance of GraphWSP over SPAGAN by an average of 20\%. This can be attributed to GraphWSP's ability to leverage perturbations in labeled data, considering more intricate interactions between data samples and optimizing the model's ability to extend label information to unlabeled datasets. Additionally, SPAGAN heavily relies on the availability of extensive feature sets. With a focus on positional embedding and limited graph element features, the model's performance is adversely affected.

A notable observation is the lower F1 score for dataset $RTD_2$. Despite its larger size compared to $RTD_1$, the represented enterprise network is in the early stages of ZT architecture deployment, characterized by a nearly fully connected network and shorter path lengths. Moreover, the assignment of criticality tags is not fully mature, impacting the model's ability to learn relationships between graph entities.

\begin{table}[ht]
\scalebox{0.9} {\begin{tabular}{lccc|ccc}
\hline
\toprule
         & \multicolumn{3}{l}{\bf Model trained by $SDT_1$} & \multicolumn{3}{l}{\bf Model trained by $RTD_1$ } \\
         \cline{2-4}\cline{4-7}
{\bf Metrics}   & $SDT_2$        & $RTD_1$     & $RTD_2$     & $SDT_1$         & $SDT_2$ & $RDT_2$
\\ \hline
Accuracy & $87.50\%$             & $85.85\%$           & $71.89\%$          & $74.18\%$  & $75.18\%$       & $47.29\%$     \\
\hline
F1       &  $83.65\%$             & $71.45\%$           &    $49.51\%$         & $68.09\%$   & $69.09\%$  & $43.14\%$          \\\hline
RoC Acc. & $88.71\%$             & $78.63\%$           & $69.86\%$             & $75.46\%$ & $76.46\%$  & $65.28\%$  \\ \hline\bottomrule       
\end{tabular}}
    \caption{\small Overview of transfer-learning evaluation of GraphWSP for feature 3 by using different datasets for training and testing.}
    \label{tab:GraphWSP evaluation}
    \vspace{-2mm}
\end{table}

\begin{table}[ht]

\scalebox{0.9} {\begin{tabular}{lccc|ccc}
\hline
\toprule
         & \multicolumn{3}{l}{\bf Model trained by $SDT_1$} & \multicolumn{3}{l}{\bf Model trained by $RTD_1$ } \\
         \cline{2-4}\cline{4-7}
{\bf Metrics}   & $SDT_2$        & $RTD_1$     & $RTD_2$     & $SDT_1$         & $SDT_2$ & $RDT_2$
\\ \hline
Accuracy & $93.80\%$             & $47.96\%$           & $56.21\%$          & $56.48\%$  & $58.15\%$ & $76.75\%$   \\ 
\hline 
F1       &  $90.75\%$             & $49.78\%$           &    $10.38\%$         & $0.0\%$   & $0.0\%$  & $0.0\%$          \\\hline
RoC Acc. & $93.20\%$             & $78.63\%$           & $8.98\%$             & $43.26\%$ & $43.36\%$  & $48.46\%$  \\ \hline\bottomrule       
\end{tabular}}
    \caption{\small Overview of transfer-learning evaluation of SPAGAN by using different datasets for training and testing.}
    \label{tab:evaluation Transfer SPAGAN}
    \vspace{-4mm}
\end{table}


\subsection{Evaluation of Pro-ZD for Risk Assessment}\label{sec:attack}

We assess the accuracy of network connection risk evaluation and classification into critical or non-critical categories (refer to \S~\ref{sec:criticality}) in two distinct settings: semi-supervised and transfer learning. Our evaluation is grounded in the context of four enterprise network datasets. We assess the model's output by comparing it to labeled datasets provided by the security teams of the contributing enterprises in this study, identifying the connections they would potentially recognize as critical.


{\bf Performance evaluation in a semi-supervised setting.} Table~\ref{tab:pro-ZD evaluation} demonstrates the accuracy of the risk assessment model (Pro-ZD). The results underscore Pro-ZD's efficacy in identifying network connections vulnerable to zero-day attacks, posing a threat to critical assets. GraphWSP's capability to discern graph structure-based features is a key contributor to the framework's success. The unique approach to calculating weighted shortest path embeddings enhances prediction interpretability without compromising precision.

\begin{table}[h]
\centering
\scalebox{1}{
\begin{tabular}{lcccc}
\hline
\toprule
         &  \multicolumn{4}{l}{\bf Pro-ZD Model} \\
        \cline{1-5}
{\bf Metrics}        & $STD_1$         & $STD_2$   & $RTD_1$ & $RTD_2$
\\ \hline
Accuracy     & $96.91\%$             & $99.005\%$ & $92.167\%$    &  $100\%$     \\
\hline
F1              & $96.12\%$             & $98.81\%$          & $92.68\%$  & $100\%$\\\hline
RoC Acc.     & $98.29\%$             & $99.60\%$ & $97.87\%$  & $100\%$  \\ \hline\bottomrule       
\end{tabular}
}
    \caption{\small Evaluation overview of Pro-ZD across four different datasets: $(STD_1)$, $(STD_2)$, $(RTD_1)$, and $(RTD_2)$.}
    \label{tab:pro-ZD evaluation}
    \vspace{-2mm}
\end{table}

In Table \ref{tab:confusion matrix}, we provide the confusion matrix for the model across the four datasets to evaluate the classification model's quality. The accurate predictions for the three larger datasets, $STD_1, STD_2$, and $RTD_2$, demonstrate the model's effectiveness. Despite the limited accuracy of the structure-based feature identified by GraphWSP for $RTD_2$, the overall performance of Pro-ZD for this dataset was was exceptionally high. However, the $RTD_1$ dataset includes some false positives and false negatives, potentially due to its relatively small size, preventing the model from capturing the intricate relationships among different graph elements during training.

\begin{table}[h]
\centering
\scalebox{0.9}{
\begin{tabular}{lcccc}
\hline
\toprule
         &  \multicolumn{4}{l}{\bf Confusion Matrix of the Four Datasets} \\
        \cline{1-5}
{\bf Datasets}        & $TP$         & $FP$   & $FN$ & $TN$
\\ \hline
STD1     & $558$             & $0$ & $16$    &  $430$     \\
\hline
STD2              & $569$             & $0$          & $16$  & $430$\\\hline
RTD1     & $194$             & $1$ & $45$  & $143$  \\ \hline
RTD2     & $2380$             & $0$ & $0$  & $1981$  \\ \hline\bottomrule       
\end{tabular}
}
    \caption{\small Evaluation overview of Pro-ZD confusion matrix across four different datasets: $STD_1, STD_2, RTD_1$ and  $RTD_2$.}
    \label{tab:confusion matrix}
    \vspace{-2mm}
\end{table}

In addition to the raw accuracy rates, we present the receiver operating characteristic curve (ROC) and area under the curve (AUC). This evaluation assesses the classifier's ability to distinguish between critical and safe edges. True positive samples represent critical instances correctly classified as critical, while false positive samples are critical instances misclassified as safe. As depicted in Figure~\ref{fig:PRO-ZDRocSemi}, the ROC curve, situated closer to the top-left corner, indicates robust discrimination of the two classes. Notably, the AUC score for larger datasets significantly surpasses that of smaller datasets, with values exceeding 0.95 versus approximately 0.8, respectively.

\begin{figure}[h]
    \centering     
    \includegraphics[width=1\columnwidth, height=0.5\columnwidth]{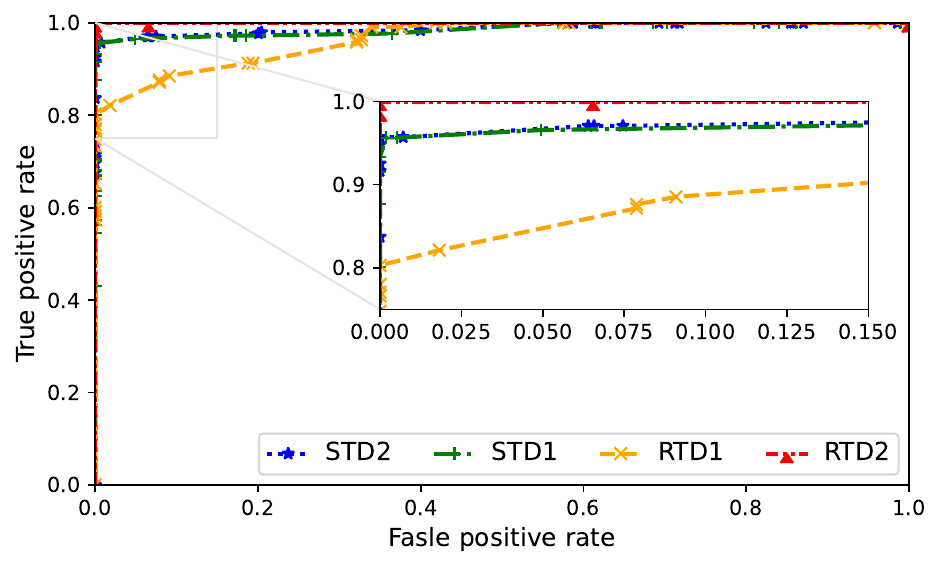}
    \caption{\small ROC curves of the PRO-ZD edge classification in the semi-supervised setting.}
    \vspace{-4mm}
    \label{fig:PRO-ZDRocSemi}
\end{figure}

Next, we examine the training dynamics, as illustrated in Figure \ref{fig:Epochs}. Empirically, we observe the convergence of our proposed method within a maximum of 40 epochs. Notably, the graph demonstrates that each dataset converges to a distinct endpoint, signifying the model's capacity to capture relationships within diverse network structures. This underscores the robustness, adaptability, and generalization capability of our proposed framework to previously unseen data.

\begin{figure}[h]
    \centering
     
    \includegraphics[width=1\columnwidth, height=0.5\columnwidth]{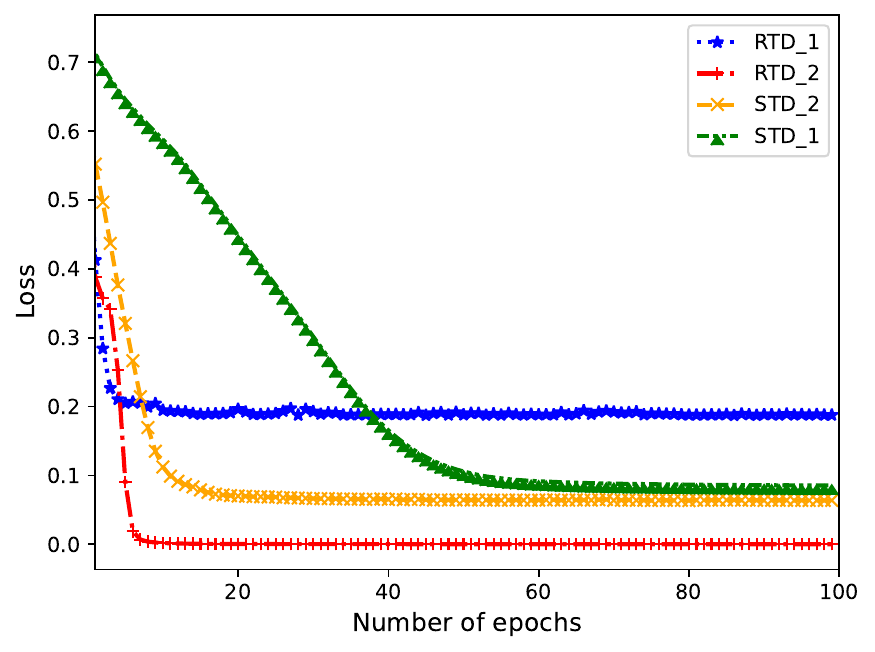}
    \caption{\small $PROD-ZD$ Cross Entropy loss convergence.}   
    \label{fig:Epochs}
    \vspace{-2mm}
\end{figure}

{\bf Performance evaluation in self-supervised setting.} To evaluate the end-to-end transferability of the proposed approach, we train the edge classifier on one dataset and evaluate its performance on different datasets. The classification accuracy results, as documented in Table~\ref{tab: evaluation transfer Pro-ZD}, affirm the inductive capabilities of Pro-ZD and its effectiveness in accurately characterizing previously unseen data.

\begin{table}[h]

\scalebox{0.88} {\begin{tabular}{lccc|ccc}
\hline
\toprule
         & \multicolumn{3}{l}{\bf Model trained by $SDT_1$} & \multicolumn{3}{l}{\bf Model trained by $RTD_1$ } \\
         \cline{2-4}\cline{4-7}
{\bf Metrics}   & $SDT_2$        & $RTD_1$     & $RTD_2$     & $SDT_1$         & $SDT_2$ & $RDT_2$
\\ \hline
Accuracy & $97.82\%$             & $88.34\%$           & $93.18\%$          & $97.82\%$  & $98.72\%$       & $91.00\%$     \\
\hline
F1       &  $97.46\%$             & $88.08\%$           &    $91.92\%$         & $97.46\%$   & $98.85\%$  & $89.02\%$          \\\hline
RoC\_Acc. & $98.99\%$             & $90.44\%$           & $100.00\%$             & $98.91\%$ & $99.48\%$  & $99.99\%$  \\ \hline\bottomrule       
\end{tabular}}
    \caption{\small Overview of transfer-learning evaluation of Pro-ZD  by using different datasets for training and testing.}
    \label{tab: evaluation transfer Pro-ZD}
    \vspace{-2mm}
\end{table}

We evaluate the ROC curve of Pro-ZD by graphing the true-positive rate (TPR) against the false-positive rate (FPR), illustrating the performance of the downstream classification task across various classification thresholds. Pro-ZD's performance in the self-supervised setting closely mirrors that of the transductive setting, with a curve positioned near the top-left corner. Optimal performance is observed around a 0.8 decision threshold, as depicted in Figure \ref{fig:ROC_transfer_settings}.

\begin{figure}[h]
    \centering
     
    \includegraphics[width=1\columnwidth, height=0.5\columnwidth]{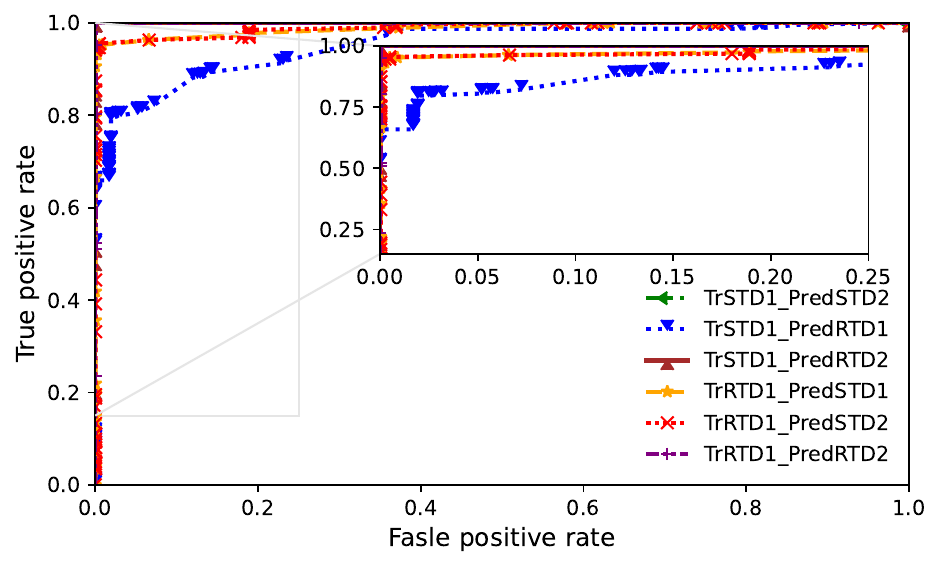}
    \caption{\small ROC curves of the {\bf $PRO-ZD$} edge classification in the transfer-learning setting.}
   
    \label{fig:ROC_transfer_settings}
    \vspace{-4mm}
\end{figure}

\section{Conclusion}
\label{sec:conclusion} 
This study represents the initial foray into GNN-based autonomous mitigation of potential zero-day attacks. Our contribution addresses existing gaps and extends the current literature by introducing a novel GNN-based approach for automated risk assessment of network connections that may expose critical assets. Additionally, we propose a framework for automated mitigation through proactive, non-person-based adjustments to network firewall rules and ZT policies, reinforcing cyber defenses before potential damage occurs. A key strength of our approach lies in its ability to identify risks beyond known vulnerabilities, allowing effective and proactive mitigation in dynamic networks where new attack vectors and sophisticated threats constantly emerge.

We introduce a novel GNN architecture, GraphWSP, designed for calculating weighted shortest path lengths. Leveraging an innovative, dynamically expandable, and transferable graph-attentional network approach, we systematically enhance the structural representation power of GNNs. GraphWSP's inductive property enables the model to leverage feature information from graph elements, generating node embeddings efficiently for previously unseen data. This self-adaptive capability ensures the model's effectiveness amid dynamic changes in network structure. Experimental results demonstrate that GraphWSP consistently outperforms the baseline SPAGAN approach across various evaluation metrics.

\bibliographystyle{IEEEtran}
\bibliography{bibl.bib}

@misc{ACSC,
author = {Australian Cyber Security Centre},
title = {Strategies to Mitigate
Cyber Security Incidents –
Mitigation Details },
year=2022,
howpublished={\url{https://www.cyber.gov.au/acsc/view-all-content/publications/strategies-mitigate-cyber-security-incidents}}
}

@inproceedings{byres,
  title={The use of attack trees in assessing vulnerabilities in SCADA systems},
  author={Byres, Eric J and Franz, Matthew and Miller, Darrin},
  booktitle={Proceedings of the international infrastructure survivability workshop},
  pages={3--10},
  year={2004}
}

@inproceedings{mcqueen,
  title={Quantitative cyber risk reduction estimation methodology for a small SCADA control system},
  author={McQueen, Miles A and Boyer, Wayne F and Flynn, Mark A and Beitel, George A},
  booktitle={Proceedings of the 39th Annual Hawaii International Conference on System Sciences},
  volume={9},
  pages={226--226},
  year={2006},
  organization={IEEE}
}

@article{wang2013,
author = {Wang, Shuzhen and Zhang, Zonghua and Kadobayashi, Youki},
title = {Exploring Attack Graph for Cost-Benefit Security Hardening},
year = {2013},
issue_date = {February 2013},
publisher = {Elsevier Advanced Technology Publications},
address = {GBR},
volume = {32},
number = {C},
issn = {0167-4048},
month = {feb},
pages = {158–169},
numpages = {12},

}

@article{AWAN2016,
title = {Identifying cyber risk hotspots: A framework for measuring temporal variance in computer network risk},
journal = {Computers \& Security},
volume = {57},
pages = {31-46},
year = {2016},
issn = {0167-4048},
author = {Malik Shahzad Kaleem Awan and Pete Burnap and Omer Rana}
}

@INPROCEEDINGS{noms2022,
title = {Towards a Zero-Trust Micro-segmentation Network Security Strategy: An Evaluation Framework},
year = {2022},
author = {Nardine Basta and Muhammad Ikram and Mohamed Ali Kaafar and Andy Walker},
booktitle={2022 IEEE/IFIP Network Operations and Management Symposium (NOMS 2022)}
}

@INPROCEEDINGS{deeplearningSecurity,
  author={Apruzzese, Giovanni and Colajanni, Michele and Ferretti, Luca and Guido, Alessandro and Marchetti, Mirco},
  booktitle={2018 10th International Conference on Cyber Conflict (CyCon)}, 
  title={On the effectiveness of machine and deep learning for cyber security}, 
  year={2018},
  volume={},
  number={}}

@article{deeplearningattacks,
author = {Fang, Xing and Xu, Maochao and Xu, Shouhuai and Zhao, Peng},
year = {2019},
month = {05},
pages = {},
title = {A deep learning framework for predicting cyber attacks rates},
volume = {2019},
journal = {EURASIP Journal on Information Security},
}

@article{Gnnintrusion,
  author    = {Wai Weng Lo and
               Siamak Layeghy and
               Mohanad Sarhan and
               Marcus Gallagher and
               Marius Portmann},
  title     = {E-GraphSAGE: {A} Graph Neural Network based Intrusion Detection System},
  journal   = {{NOMS}},
  year      = {2022}
}

@article{GnnAnomaly,
author = {Protogerou, Aikaterini and Papadopoulos, Stavros and Drosou, Anastasios and Tzovaras, Dimitrios and Refanidis, Ioannis},
year = {2021},
month = {03},
pages = {},
title = {A graph neural network method for distributed anomaly detection in IoT},
volume = {12},
journal = {Evolving Systems}
}

@article{GnnAnomaly2,
author = {Wu, Yulei and Dai, Hong-Ning and Tang, Haina},
year = {2021},
month = {07},
pages = {1-1},
title = {Graph Neural Networks for Anomaly Detection in Industrial Internet of Things},
volume = {PP},
journal = {IEEE Internet of Things Journal},
}

@inproceedings{Gnnvulnerability,
 author = {Zhou, Yaqin and Liu, Shangqing and Siow, Jingkai and Du, Xiaoning and Liu, Yang},
 booktitle = {Advances in Neural Information Processing Systems},
 editor = {H. Wallach and H. Larochelle and A. Beygelzimer and F. d\textquotesingle Alch\'{e}-Buc and E. Fox and R. Garnett},
 pages = {},
 publisher = {Curran Associates, Inc.},
 title = {Devign: Effective Vulnerability Identification by Learning Comprehensive Program Semantics via Graph Neural Networks},
 volume = {32},
 year = {2019}
}

@ARTICLE{Gnnvulnerability2,
  author={Wang, Huanting and Ye, Guixin and Tang, Zhanyong and Tan, Shin Hwei and Huang, Songfang and Fang, Dingyi and Feng, Yansong and Bian, Lizhong and Wang, Zheng},
  journal={IEEE Transactions on Information Forensics and Security}, 
  title={Combining Graph-Based Learning With Automated Data Collection for Code Vulnerability Detection}, 
  year={2021},
  volume={16},
  number={}
  }

@article{Gnnbots,
title = {Multi-attributed heterogeneous graph convolutional network for bot detection},
journal = {Information Sciences},
volume = {537},
pages = {380-393},
year = {2020},
issn = {0020-0255},
author = {Jun Zhao and Xudong Liu and Qiben Yan and Bo Li and Minglai Shao and Hao Peng}

}

@inproceedings{Gnnmalware,
  title     = {Heterogeneous Graph Matching Networks for Unknown Malware Detection},
  author    = {Wang, Shen and Chen, Zhengzhang and Yu, Xiao and Li, Ding and Ni, Jingchao and Tang, Lu-An and Gui, Jiaping and Li, Zhichun and Chen, Haifeng and Yu, Philip S.},
  booktitle = {Proceedings of the Twenty-Eighth International Joint Conference on Artificial Intelligence, {IJCAI-19}},
  publisher = {International Joint Conferences on Artificial Intelligence Organization},             
  year      = {2019},
  month     = {7}
}

@article{semiKipf,
  author    = {Thomas N. Kipf and
               Max Welling},
  title     = {Semi-Supervised Classification with Graph Convolutional Networks},
  journal   = {CoRR},
  volume    = {abs/1609.02907},
  year      = {2016}
}

@misc{kipf2,
  doi = {10.48550/ARXIV.1706.02263},
  url = {https://arxiv.org/abs/1706.02263},
  author = {Berg, Rianne van den and Kipf, Thomas N. and Welling, Max},
  title = {Graph Convolutional Matrix Completion},
  publisher = {arXiv},
  year = {2017},
}

@book{DL_cyber,
author = {Alazab, Mamoun and Tang, MingJian},
title = {Deep Learning Applications for Cyber Security},
year = {2019},
isbn = {3030130568},
publisher = {Springer Publishing Company, Incorporated},
edition = {1st}
}

@INPROCEEDINGS{ZDATT,
author={Sun, Xiaoyan and Dai, Jun and Liu, Peng and Singhal, Anoop and Yen, John},
booktitle={2016 IEEE Conference on Communications and Network Security (CNS)}, 
title={Towards probabilistic identification of zero-day attack paths}, 
year={2016},
volume={},
number={},
pages={64-72},
}

@InProceedings{patrol,
author="Dai, Jun
and Sun, Xiaoyan
and Liu, Peng",
editor="Crampton, Jason
and Jajodia, Sushil
and Mayes, Keith",
title="Patrol: Revealing Zero-Day Attack Paths through Network-Wide System Object Dependencies",
booktitle="Computer Security -- ESORICS 2013",
}

@InProceedings{ZD1,
author="Ye, Ziwei
and Guo, Yuanbo
and Ju, Ankang",
editor="Sun, Xingming
and Pan, Zhaoqing
and Bertino, Elisa",
title="Zero-Day Vulnerability Risk Assessment and Attack Path Analysis Using Security Metric",
booktitle="Artificial Intelligence and Security",
year="2019",
publisher="Springer International Publishing",
address="Cham",
pages="266--278",
}

@ARTICLE{k-zero,
author={Wang, Lingyu and Jajodia, Sushil and Singhal, Anoop and Cheng, Pengsu and Noel, Steven},
journal={IEEE Transactions on Dependable and Secure Computing}, 
title={k-Zero Day Safety: A Network Security Metric for Measuring the Risk of Unknown Vulnerabilities}, 
year={2014},
volume={11},
number={1},
pages={30-44},
}

@inproceedings{distanceEncoding,
author = {Li, Pan and Wang, Yanbang and Wang, Hongwei and Leskovec, Jure},
title = {Distance Encoding: Design Provably More Powerful Neural Networks for Graph Representation Learning},
year = {2020},
publisher = {Curran Associates Inc.},
articleno = {375},
numpages = {14}
}

@inproceedings{Spagan,
  title={SPAGAN: Shortest Path Graph Attention Network},
  author={Yiding Yang and Xinchao Wang and Mingli Song and Junsong Yuan and Dacheng Tao},
  booktitle={International Joint Conference on Artificial Intelligence},
  year={2019}
}

@inproceedings{PGNN,
  author    = {Jiaxuan You and
               Rex Ying and
               Jure Leskovec},
  title     = {Position-aware Graph Neural Networks},
  booktitle = {{ICML}},
  year      = {2019}
}

@article{dl_vul1,
author = {Ghaffarian, Seyed Mohammad and Shahriari, Hamid Reza},
title = {Software Vulnerability Analysis and Discovery Using Machine-Learning and Data-Mining Techniques: A Survey},
year = {2017},
issue_date = {July 2018},
publisher = {Association for Computing Machinery},
address = {New York, NY, USA},
volume = {50},
number = {4},
journal = {ACM Comput. Surv.},
month = {aug},
articleno = {56},
numpages = {36},
}

@Article{dl_vul2,
AUTHOR = {Nikoloudakis, Yannis and Kefaloukos, Ioannis and Klados, Stylianos and Panagiotakis, Spyros and Pallis, Evangelos and Skianis, Charalabos and Markakis, Evangelos K.},
TITLE = {Towards a Machine Learning Based Situational Awareness Framework for Cybersecurity: An SDN Implementation},
JOURNAL = {Sensors},
VOLUME = {21},
YEAR = {2021},
NUMBER = {14},
ARTICLE-NUMBER = {4939}
}

@ARTICLE{dl_vul3,
  author={Zolanvari, Maede and Teixeira, Marcio A. and Gupta, Lav and Khan, Khaled M. and Jain, Raj},
  journal={IEEE Internet of Things Journal}, 
  title={Machine Learning-Based Network Vulnerability Analysis of Industrial Internet of Things}, 
  year={2019},
  volume={6}
}

@article{AG_ZD,
author = {Kumar, Vikash and Sinha, Ditipriya},
year = {2021},
month = {05},
pages = {},
title = {A robust intelligent zero-day cyber-attack detection technique},
volume = {7},
journal = {Complex \& Intelligent Systems}
}

@inproceedings{deepwalk,
author = {Perozzi, Bryan and Al-Rfou, Rami and Skiena, Steven},
title = {DeepWalk: Online Learning of Social Representations},
year = {2014},
booktitle = {Proceedings of the 20th ACM SIGKDD International Conference on Knowledge Discovery\& Data Mining}
}

@inproceedings{GNNWEAK1,
  author    = {Keyulu Xu and
               Weihua Hu and
               Jure Leskovec and
               Stefanie Jegelka},
  title     = {How Powerful are Graph Neural Networks?},
  booktitle = {7th International Conference on Learning Representations, {ICLR} 2019,
               New Orleans, LA, USA, 2019},
  year      = {2019},

}

@article{assetValuation,
  title={IT Asset Valuation , Risk Assessment and Control Implementation Model},
  author={Shemlse Gebremedhin Kassa},
  year={2017},
  journal = {ISACA},
  volume = {3},
}

@book{app_criticality,
title = {Managing risk in information systems },
year = {2014 - 2015},
author = {Gibson, Darril},
booktitle = {Managing risk in information systems},
edition = {2},
}

@article{SP_complexity,
author = {Lov\'{a}Sz, L\'{a}Szl\'{o}},
title = {Review of the Book by Alexander Schrijver: Combinatorial Optimization: Polyhedra and Efficiency},
year = {2005},
volume = {33},
journal = {Oper. Res. Lett.},
}

@article{SPGNN,
  title={SPGNN-API: A Transferable Graph Neural Network for Attack Paths Identification and Autonomous Mitigation},
  author={Houssem Jmal and Firas Hmida and Nardine Basta and Muhammad Ikram and Mohamed Ali K{\^a}afar and Andy Walker},
  journal={IEEE Transactions on Information Forensics and Security},
  year={2023},
  volume={19},
  pages={1601-1613},
}

@inproceedings{GAT,
title={Graph Attention Networks},
author={Petar Veličković and Guillem Cucurull and Arantxa Casanova and Adriana Romero and Pietro Liò and Yoshua Bengio},
booktitle={International Conference on Learning Representations},
year={2018}
}

@inproceedings{SL,
  author    = {Felix Wu and
               Amauri H. Souza Jr. and
               Tianyi Zhang and
               Christopher Fifty and
               Tao Yu and
               Kilian Q. Weinberger},
  title     = {Simplifying Graph Convolutional Networks},
  booktitle = {Proceedings of the 36th International Conference on Machine Learning, {ICML}},
year={2018}


}

@inproceedings{multihead,
author = {Vaswani, Ashish and Shazeer, Noam and Parmar, Niki and Uszkoreit, Jakob and Jones, Llion and Gomez, Aidan N. and Kaiser, \L{}ukasz and Polosukhin, Illia},
title = {Attention is All You Need},
year = {2017},
publisher = {Curran Associates Inc.},
booktitle = {{NIPS}}
}

@INPROCEEDINGS{crimes,
  author={Sharma, Vinita and Manocha, Tanu and Garg, Seema and Sharma, Saatwik and Garg, Anshita and Sharma, Ritu},
  booktitle={{ICIPTM}}, 
  title={Growth of Cyber-crimes in Society 4.0}, 
  year={2023}
}

@inproceedings{Node2vec,
author = {Grover, Aditya and Leskovec, Jure},
title = {Node2vec: Scalable Feature Learning for Networks},
year = {2016},
booktitle = {Proceedings of the 22nd ACM SIGKDD International Conference on Knowledge Discovery and Data Mining},
}

 
\vspace{-30pt}
\begin{IEEEbiography}[{\includegraphics[width=1in,height=1.25in,clip,keepaspectratio]{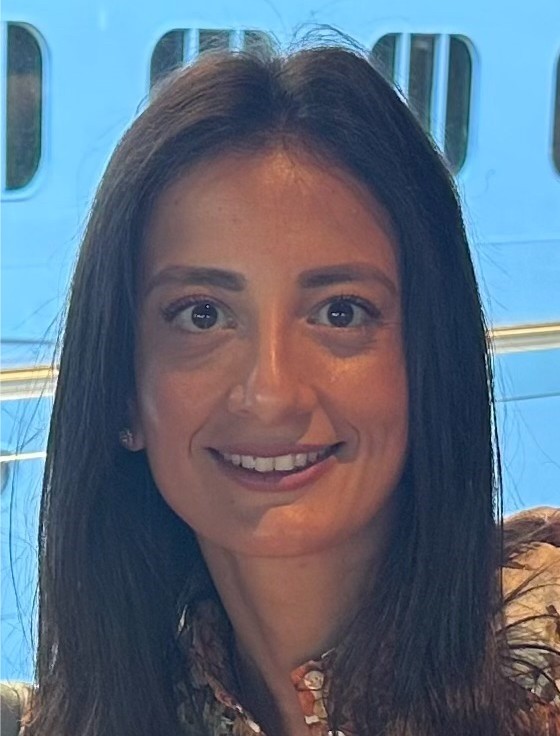}}]{Nardine Basta} holds a Master's degree from Johannes Kepler University in Austria. She received her Ph.D. in 2019 from the University of Ulm, Germany. She is currently contributing to cutting-edge research as a postdoctoral researcher at the Macquarie University Cybersecurity Hub in Sydney, Australia. Her research interests lie in the intersection of network security and neural networks. She specializes in graph neural networks and large language models, contributing to advancements in these critical domains.
\end{IEEEbiography}
\vskip 0pt plus -1fil 
\begin{IEEEbiography}[{\includegraphics[width=1in,height=1.25in,clip,keepaspectratio]{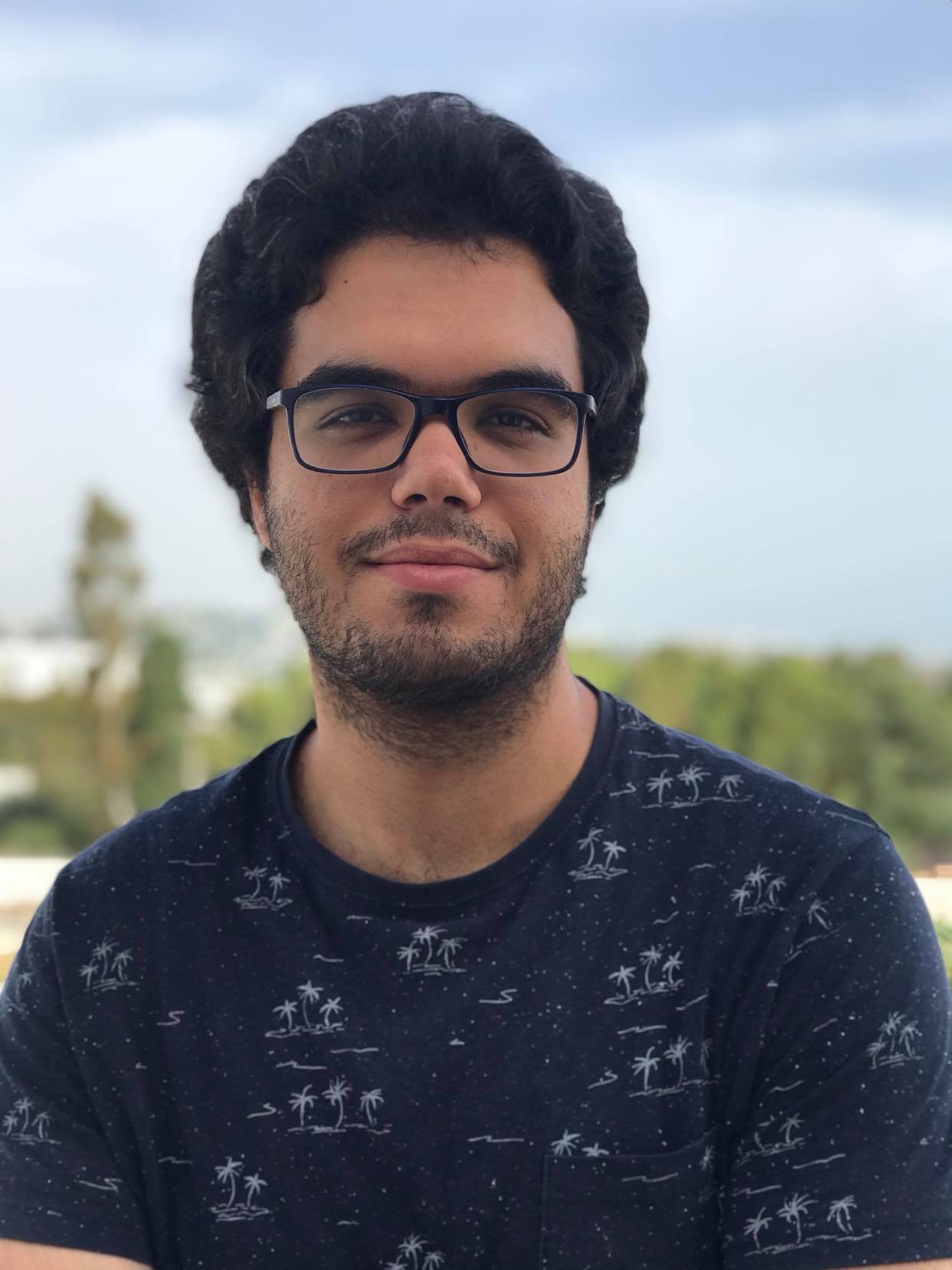}}]{Houssem Jmal} earned his engineering degree with a specialization in signals and systems from Ecole Polytechnique de Tunisie in 2023. He complemented his studies with hands-on experience through internships, notably at the Macquarie University Cyber Security Hub. Currently pursuing a Ph.D. at Nantes Université within LS2N lab (STACK team). Houssem's doctoral investigations revolve around the intricacies of Federated Learning for Enhancing the Security and Privacy of Decentralized and Distributed Systems, a critical facet of the European Project Di4SPDS. His research interest is primarily focused on AI, with a specific tendency toward its practical applications in fortifying cybersecurity.
\end{IEEEbiography}
\vskip 0pt plus -1fil 
\begin{IEEEbiography}[{\includegraphics[width=1in,height=1.25in,clip,keepaspectratio]{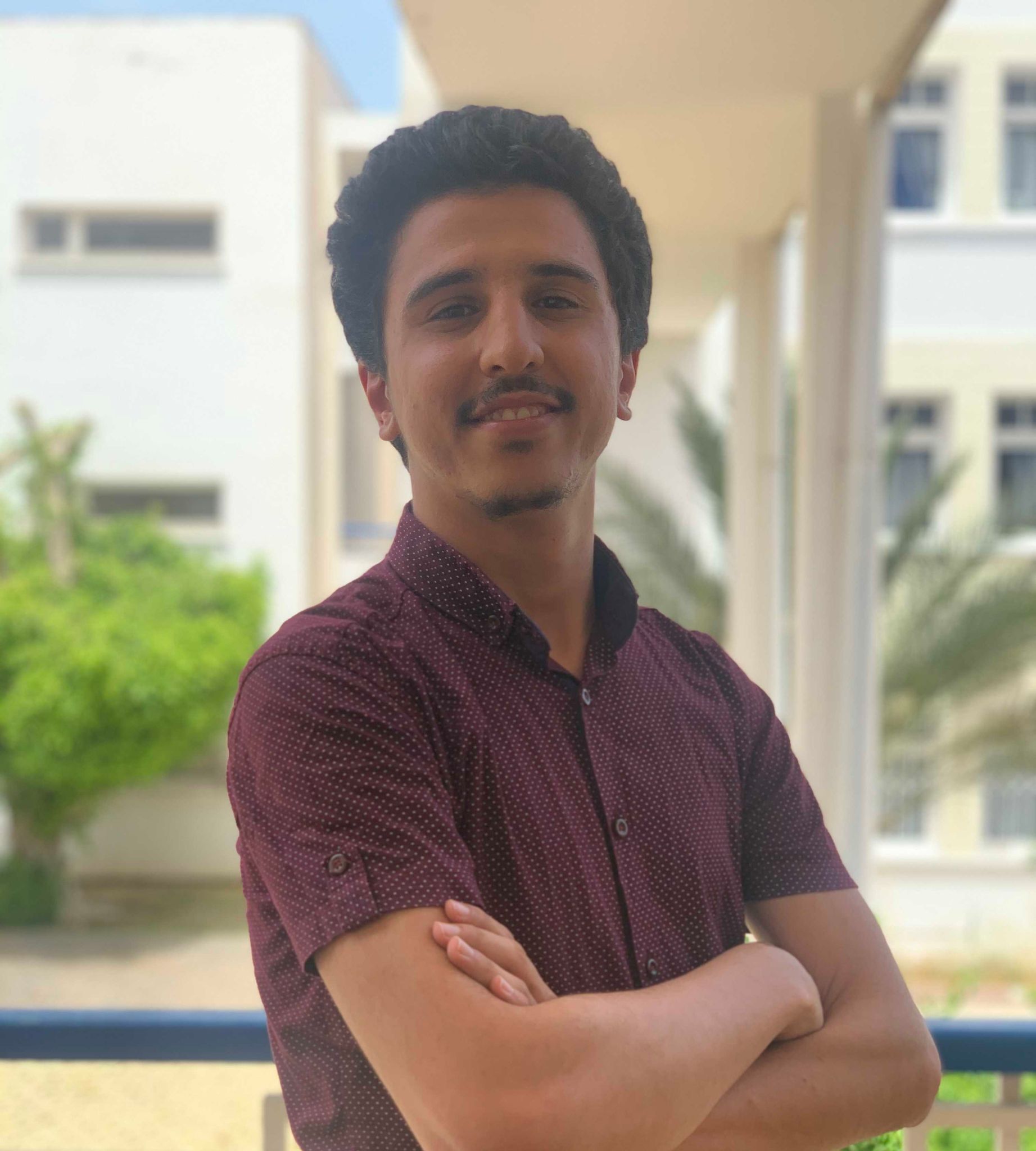}}]{Firas Ben Hmida} earned his engineering degree with a specialization in signals and systems from Ecole Polytechnique de Tunisie in 2023. He complemented his studies with hands-on experience through internships, notably at the Macquarie University Cyber Security Hub. Currently pursuing a Ph.D. at the University of Michigan Dearborn. His research area is focused on adversarial machine learning, robustness, AI Trustworthiness, and provenance capture and analysis as the basis for a robust cybersecurity defense strategy for AI models.
\end{IEEEbiography}
\vskip 0pt plus -1fil 
\begin{IEEEbiography}[{\includegraphics[width=1in,height=1.25in,clip,keepaspectratio]{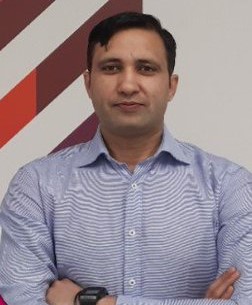}}]{Muhammad Ikram} received the Ph.D. degree from the University of New South Wales, in 2018. He was a joint Postdoctoral Research Fellow with the University of Michigan, USA, and Cyber Security Hub Macquarie University. He was a Visiting Scientist with CSIRO, from 2018 to 2020. He is currently a Lecturer with the Department of Computing, Macquarie University. Over the last couple of years, he has been thinking and working to build techniques leveraging machine-learning algorithms to fight against security and privacy issues and fraud detection targeting online services. He has several publications in prestigious measurement, security, and privacy conferences, such as USENIX Security Symposium, NDSS, PETS, and IMC, and journals, such as British Medical Journal and ACM Transactions on Privacy and Security. His research contributions are featured by several news and media outlets reaching millions of audiences worldwide. His current research interests include large-scale measurements, analytics, and analyzing security and privacy issues in web and mobile platforms.
\end{IEEEbiography}
\vskip 0pt plus -1fil 
\begin{IEEEbiography}
[{\includegraphics[width=1in,height=1.25in,clip,keepaspectratio]{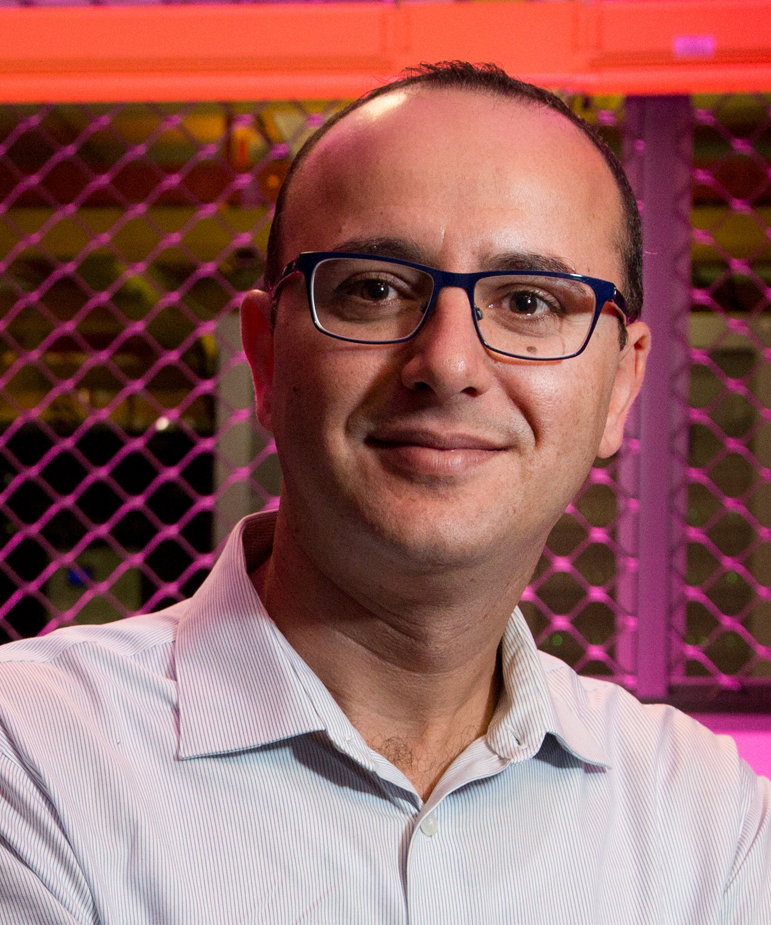}}]{Mohamed Ali (Dali) Kaafar} received the Ph.D. degree from the University of Nice Sophia Antipolis and INRIA, France, where he pioneered research in the security of internet coordinate systems. He is currently a Professor at the Faculty of Science and Engineering, Macquarie University, and the Executive Director of the Macquarie University Cyber Security Hub. He is also the Founder of the Information Security and Privacy Group, CSIRO Data61. Prior to that, he was the Research Leader of the Data Privacy and Mobile Systems Group, NICTA, and a Senior Principal Researcher at the INRIA, the French research institution of computer science and automation. He published over 300 scientific peer-reviewed articles with several repetitive publications in the prestigious IEEE Symposium on Security and Privacy (IEEE S\&P), ACM SIGCOMM, WWW, NDSS, and PETS. He received several awards, including the INRIA Excellence of Research National Award, and the Andreas Pfitzman Award from the Privacy Enhancing Technologies Symposium in 2011. In 2019, he has also been awarded the prestigious and selective Chinese Academy of Sciences President’s Professorial Fellowship Award. He is an Associate Editor of the IEEE TRANSACTIONS ON INFORMATION FORENSICS AND SECURITY and serves on the Editorial Board for the Journal on Privacy Enhancing Technologies.
\end{IEEEbiography}

\end{document}